\documentclass[aip,jcp,amsmath,amssymb,reprint]{revtex4-1}

\usepackage{graphicx}
\usepackage{dcolumn}
\usepackage{bm}
\usepackage{mathrsfs}
\usepackage[dvipsnames]{xcolor}
\usepackage[normalem]{ulem}
\usepackage{caption}
\usepackage[textfont=rm]{subcaption}

\begin{document}
\title{Connecting real glasses to mean-field models}

\author{Ujjwal Kumar Nandi}
\affiliation{\textit{Polymer Science and Engineering Division, CSIR-National Chemical Laboratory, Pune-411008, India}}
\affiliation{\textit{Academy of Scientific and Innovative Research (AcSIR), Ghaziabad 201002, India}}

\author{Walter Kob}
\email{walter.kob@umontpellier.fr}

\affiliation{\textit{Laboratoire Charles Coulomb and CNRS, University of Montpellier, Montpellier F-34095, France}}

\author{Sarika Maitra Bhattacharyya}
\email{mb.sarika@ncl.res.in}
\affiliation{\textit{Polymer Science and Engineering Division, CSIR-National Chemical Laboratory, Pune-411008, India}}
\affiliation{\textit{Academy of Scientific and Innovative Research (AcSIR), Ghaziabad 201002, India}}

\date{\today}

\begin{abstract}
We propose a novel model for a glass-forming liquid which allows to switch in a continuous manner from a standard three-dimensional liquid to a fully connected mean-field model. This is achieved by introducing $k$ additional particle-particle interactions which thus augments the effective number of neighbors of each particle. Our computer simulations of this system show that the structure of the liquid does not change with the introduction of these pseudo neighbours and by means of analytical calculations, we determine the structural properties related to these additional neighbors. We show that the relaxation dynamics of the system slows down very quickly with increasing $k$ and that the onset and the mode-coupling temperatures increase. The systems with high values of $k$ follow the MCT power law  behaviour for a larger temperature range compared to the ones with lower values of $k$. The dynamic susceptibility indicates that the dynamic heterogeneity decreases with increasing $k$ whereas the non-Gaussian parameter is independent of it. Thus we conclude that with the increase in the number of pseudo neighbours the system becomes more mean-field like. By comparing our results with previous studies on mean-field like system we come to the conclusion that the details of how the mean-field limit is approached are important since they can lead to different dynamical behavior in this limit.

\end{abstract}

\maketitle

\section{Introduction:}
The details of the relaxation dynamics of glassy system and the properties of the glass has been and continues to be in the focus of an intense research activity~\cite{binder_kob_book}. These investigations are motivated by the fact that glasses are not only important for many daily and technological applications but are also an intellectual challenge for fundamental studies since so far there is no theoretical framework that is able to give a satisfactory description of the unusual properties of glassy systems and glasses. Although there are sophisticated mean-field theories, like the mode-coupling theory (MCT) of the glass transition \cite{gotze,gotze_book,janssen_frontphys_2018,das_book}, or the random first order transition theory \cite{wolynes,xia_woly_pnas,wolynes_lubchenko}, that are able to give in some cases a surprisingly good description of real glass former \cite{kob-andersen,kob-andersen-II,gleim,sarika_PNAS,manoj_ctrw,chong}, these approaches still have many flaws since they fail to give a reliable description of many features of glass-forming systems opening thus the door to other approaches that attempt to describe glassy systems \cite{szamel,unravel,manoj-prl,manoj_softness,chandler_prl,chandler_jcp}. Note that these theories are mean-field in nature, whereas the experiments and computer simulation studies are three or lower dimensional systems. Moreover, it has been found that MCT, although expected to be mean-field in nature, does not become exact even at high dimensions \cite{ikeda-kuni,schmid-schilling}, a flaw which might, however, be related to the approximations used to describe the structure of the liquid in high dimensions. Thus it is important to understand how these theories are connected to real glass-forming systems and how the properties change as the mean-field character of the system is modified. To establish such a connection it is useful to study systems whereby varying a parameter one can go from $d$ dimensional system to mean-field (MF) system. In the past various possibilities have been proposed to take this limit, see Ref.~\onlinecite{mari-kurchan} for an overview, but most of them do have some drawbacks that prevent to reach a solid understanding how three-dimensional (3d) and MF systems are related to each other~\cite{mari-kurchan}.

One interesting model that allows approaching the MF limit in a continuous manner has been proposed by Mari and Kurchan (MK)~\cite{mari-kurchan}. The MK-model is a hard-sphere system in which the interaction range between two particles $i$ and $j$ is a random variable with a variance that allows switching  from a standard three-dimensional system to MF like system. For this model, it is found that with increasing interaction range the Stokes-Einstein relation holds down to lower temperatures and that the dynamic heterogeneity of  the system, measured by the four-point susceptibility and non-Gaussian parameter, decreases. The increase in interaction range also makes the system follow MCT like behaviour for a larger range in temperature. Although all these results indicate that the MK model can indeed be used to study the transition from 3d to MF, there are certain features of the model that are disturbing. First of all, the structural properties of the system becomes very different from the one of a normal liquid if the MF limit is approached in that, e.g.,~the radial distribution function becomes gas-like. Related to this is the fact that the three-point correlation functions vanish. As a consequence one looses the property that nearest neighbors can cage a tagged particle, a notion that is fundamental for the slowing down of the dynamics in real glass-forming systems~\cite{binder_kob_book}. Secondly, the maximum attainable packing fraction diverges in the MF limit, a behavior that is very different from the one found in finite dimensions. Some of these oddities are avoided if one considers models on a lattice~\cite{berthier_pre_86_031502_(2012)}.
However, lattice models, notably kinetic Ising models with non-conserved particle density, do have the drawback that it is not obvious to what extent their relaxation dynamics is related to any off-lattice systems. As a consequence one has to be cautious when applying results from lattice models to describe the dynamics of real systems.

Another approach to connect the properties of 3d systems with the MF behavior has been proposed in a series of papers by Miyazaki and coworkers who have studied the properties of the Gaussian-Core-Model (GCM)~\cite{kuni-pre,kuni-jcp,kuni-prl}.  Due to the long interaction range, each particle has a large number of neighbours, and hence the system can be expected to be MF like. These authors showed that compared to the (short-ranged) Kob-Andersen (KA) model \cite{kob-andersen}, in the GCM the Stokes-Einstein relation is followed till a lower temperature regime and that the relaxation dynamics shows a qualitatively better agreement with the MCT predictions~\cite{kuni-jcp}. Furthermore, it was found  that the GCM shows less dynamic fluctuation and that activated processes are suppressed~\cite{kuni-pre}, in agreement with recent studies of the thermodynamic properties of this system~\cite{manoj-gcm}. 

A further possibility to connect the properties of low dimensional systems with the MF predictions is to consider systems with increasingly higher dimensions.  Sengupta {\it et al.} have studied the properties of some standard glass formers in 2, 3, and 4 dimensions and found that with increasing dimensionality the breakdown of the Stokes-Einstein relation becomes less pronounced and that the dynamical heterogeneity decrease~\cite{sastry-SE}. Charbonneau {\it et al.} have studied systems up to 6 dimensions and found that the shape of the cage does not become Gaussian-like, as expected from MF~\cite{patrick}, showing that the approach to this limit might be more complex than expected.  

In the present paper we introduce a simple approach that allows crossing over in a continuous manner from a normal 3d liquid to a MF system. In practice we do this by increasing for each particle the number of particles it can interact with, thus increasing the effective interaction of the particle with the rest of the system. In contrast to the studies discussed above, our method does not modify in a significant manner the local structure of the liquid even when the MF limit is reached, i.e.~the structure is always similar to the one of the 3d system. So this allows us to study how increasing connectivity affects the relaxation dynamics, without modifying in a noticeable manner the structure, and hence to probe the dynamics upon approaching the MF limit.

The rest of the paper is organized as follows: The system and simulation details are described in Sec.~\ref{sec:II}. In Sec.~\ref{sec:III}, we  present the result while in Sec.~\ref{sec:IV} we summarize and conclude.

\section{Details of system and simulations}
\label{sec:II}

As mentioned in the Introduction, our system is given by $N$ particles that interact with each other via a standard short-range potential. In addition, each particle interacts also with ``pseudo neighbors'', i.e.~particles that are not necessarily close in space.
Hence the total interaction potential of the system is given by

\begin{eqnarray}
U_{\rm tot}(r_{1},...r_{N})&=&\sum_{i=1}^{N}\sum_{j>i}^{N}u(r_{ij})+\sum_{i=1}^{N}\sum_{j=1}^{k}u^{\rm pseudo}(r_{ij}) \; \;
\label{eq1}\\
&=&U+U^{\rm pseudo}_{k} \qquad .
\label{eq2}
\end{eqnarray}

\noindent
The first term on the right-hand side is the regular interaction between particles while the second term is the interaction each particle has with its pseudo neighbours. Here we consider the case that the regular interaction describes  a binary Lennard-Jones (LJ) system, with 80\% of the particles of type A and 20\% of the particles of type B. Thus the interaction between the particles $i$ and $j$ is given by

\begin{equation}
u(r_{ij})=4\epsilon_{ij}\Big[\Big(\frac{\sigma_{ij}}{r_{ij}}\Big)^{12}-\Big(\frac{\sigma_{ij}}{r_{ij}}\Big)^6\Big] \quad,
\end{equation}

\noindent
where $r_{ij}$ is the distance between the particles, $\sigma_{ij}$ is the effective diameter of the particle and $\epsilon_{ij}$ is the interaction strength. We use $\sigma_{AA}$ and $\epsilon_{AA}$ as the unit of length and energy, setting the Boltzmann constant $k_B=1$. The values of the other parameters are given in Ref.~\citenum{kob-andersen}, i.e.~$\sigma_{AB}=0.8$,  $\sigma_{BB}$=0.88, $\epsilon_{AB}$=1.5, and $\epsilon_{BB}$=0.5, a choice which makes this binary system to be a good glass-former. This potential is cut and shifted at $r_c=2.5\sigma_{ij}$. The masses are $m_A=m_B=1$ and time is expressed in units of $\sqrt{m_A \sigma^2_{AA}/\epsilon_{AA}}$.

The interaction potential with the pseudo neighbours is modelled in terms of a modified LJ potential,

\begin{figure}[tb]
\includegraphics[width=0.45\textwidth,trim = {0 0cm 0 0.1cm},clip ]{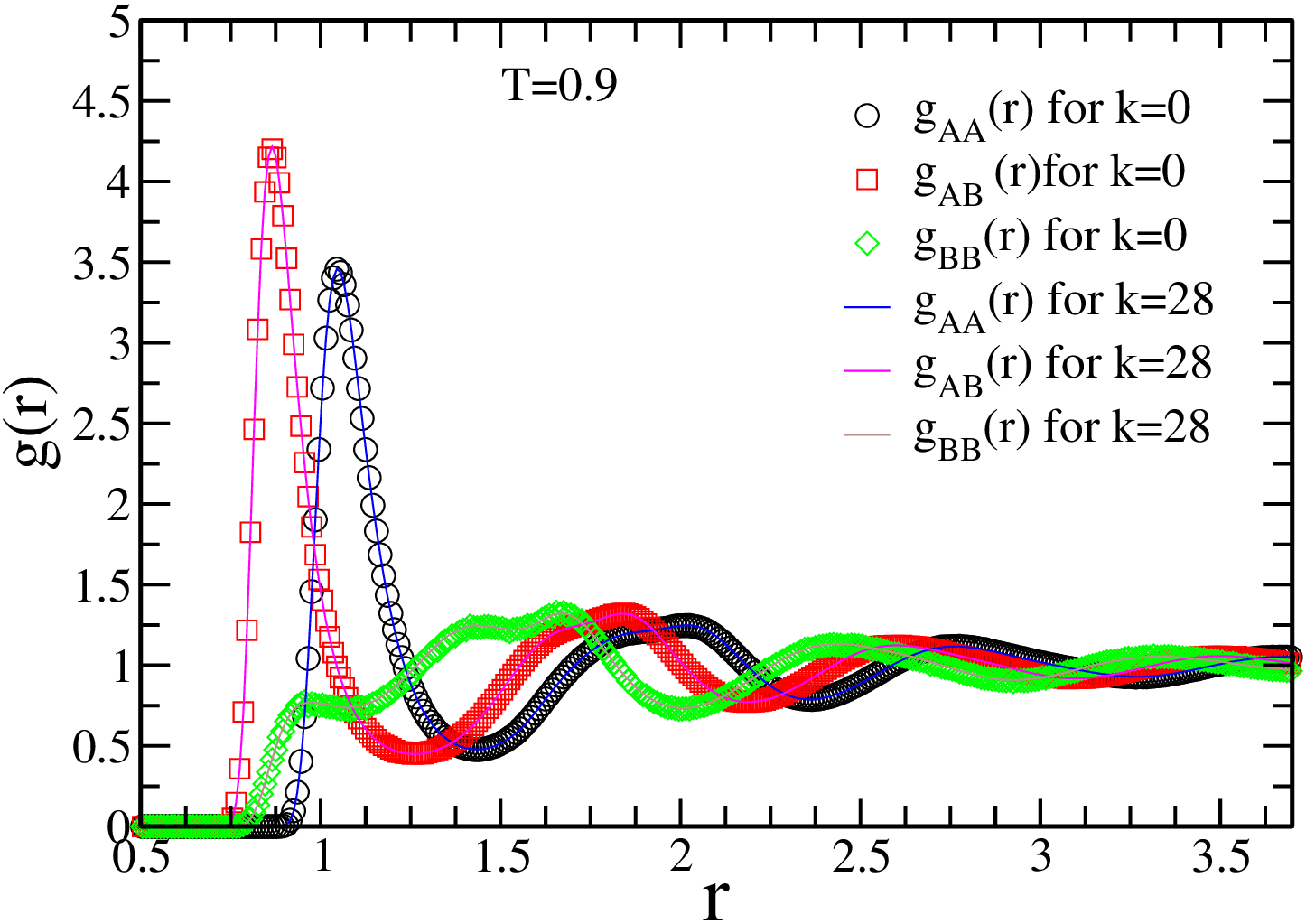}
\caption{The partial radial distribution functions for $k=0$ and $k=28$ at $T=0.9$. The structure remains invariant under the introduction of the pseudo neighbours.
}
\label{fig1_gr_regular}
\end{figure}

\begin{eqnarray}
u^{\rm pseudo}(r_{ij})&=&u(r_{ij}-L_{ij}) \\
&=&4\epsilon_{ij}\Big[\Big(\frac{\sigma_{ij}}{r_{ij}-L_{ij}}\Big)^{12}-\Big(\frac{\sigma_{ij}}{r_{ij}-L_{ij}}\Big)^6\Big] \quad ,
\end{eqnarray}

\noindent
where $L_{ij}$ is a random variable defined below. In our simulations we impose the restriction that any two particles interact either via $u(r_{ij})$ or via $u^{\rm pseudo}(r_{ij})$. This condition determines how for a given configuration equilibrated with the potential $u$ the pseudo neighbors and the values $L_{ij}$ are chosen: Taking this configurations we select for each particle, $i$,
$k$ random numbers $L_{ij}$ in the range $r_c \leq L_{ij} \leq L_{\rm max}$, where $L_{\rm max} \leq L_{\rm box}/2-r_c$, with $L_{\rm box}$ the size of the simulation box. (The distribution of these random variables will be denoted by $\mathscr{P}(L_{ij})$ and in the following, we will consider the case that the distribution is uniform.) Subsequently we choose $k$ distinct  particles $j$ with $r_{ij}>r_c$ and use the $L_{ij}$ to fix {\it permanently} the interaction between particles $i$ and $j$. This procedure thus makes that each particle $i$ interacts not only with the particles that are within the cutoff distance but in addition to $k$ particles that can be far away. Note that once the particle $j$ is chosen as a pseudo neighbour of particle $i$, automatically particle $i$ becomes a pseudo neighbour of particle $j$. The system, as defined here, can then be simulated using a standard simulation algorithms.

The molecular dynamics (MD) simulation have been done using $N=2744$ particles. We have performed constant volume, constant temperature simulations (velocity rescaling) at density $\rho=1.2$, thus $L_{\rm box}=13.1745$, using a time integration step of $\Delta t=0.005$. For $L_{\rm max}$ we have taken 4.0, slightly below the maximum value of 4.09. We have simulated four different systems with the number of pseudo neighbours, $k=0,4,12,$ and 28.

\section{Results}
\label{sec:III}

\subsection{Structure of the liquid}

\begin{figure}[!htb]
\includegraphics[width=0.45\textwidth,trim = {0 0cm 0 0.1cm},clip]{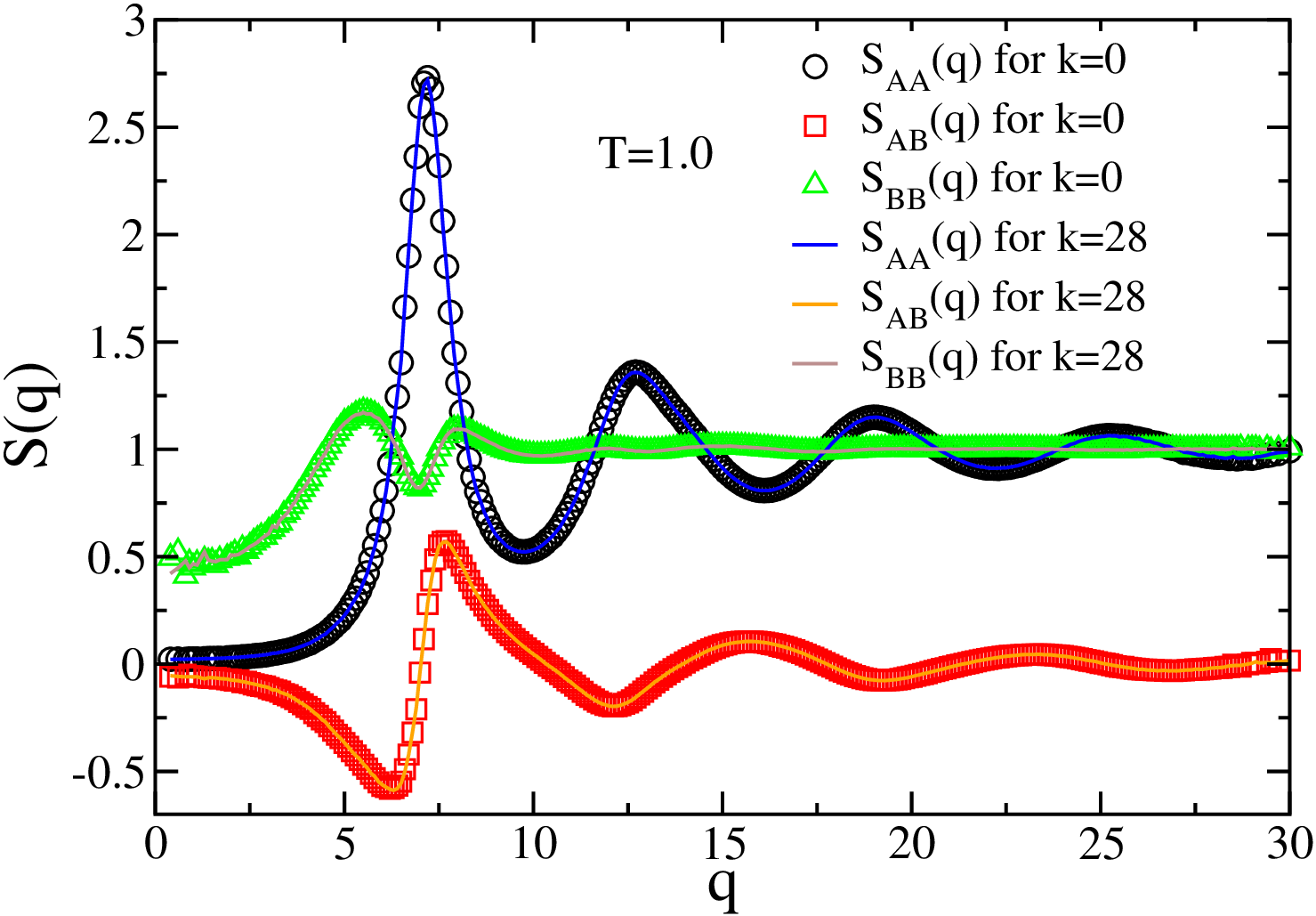}
\caption{The partial structure factors for $k=0$ and $k=28$ at $T=1.0$. Similar to what we have obtained in the radial distribution function, the structure remains invariant under the introduction of the pseudo neighbours.
}
\label{fig2_sq_regular}
\end{figure}

To start, we discuss the effect of the pseudo neighbours on the structure of the liquid. In Fig.~\ref{fig1_gr_regular} we show the three partial radial distribution function, $g_{\alpha\beta}(r)$ with $\alpha, \beta \in \{A,B \}$~\cite{hansen_mcdonald_86},  for the $k=0$ and the $k=28$ systems. The temperature is $T=0.9$, which for the $k=0$ system is slightly above the onset temperature, see Ref.~\citenum{kob-andersen}, while for the $k=28$ system it corresponds to a state at which the system is already rather viscous (see below). The graph shows that the radial distribution functions for the two systems overlap perfectly well, i.e.~the structure is independent of $k$ for this value of $k$. Thus this indicates that the interactions due to the pseudo neighbours do not affect the local structure of the system, one of the reasons for our choice of the interactions of the model.

To probe whether the structure of the liquid on a large scale is influenced by the introduction of the pseudo neighbors we have calculated the partial static structure factors and show them in Fig.~\ref{fig2_sq_regular} for the case of $k=0$ and $k=28$. Since the two sets of curves match each other perfectly well, we can conclude that also the large scale structure is not influenced by the additional neighbors.

\subsection{Static properties of the pseudo neighbors}

In this subsection, we characterize some of the structural properties of the pseudo neighbors with respect to a tagged particle.

To start, we first calculate the probability $P_L$ that a given pseudo neighbor $j$ interacts with the tagged particle $i$, where $L=L_{ij}$. Neglecting the indirect interactions (via the direct neighbors) between the tagged particle and the pseudo neighbor one can express $P_L$ as

\begin{equation}
P_{L}=\frac{\int_{V_{\rm acc}} d{\bf r} \, e^{-\beta u(r-L)}y(r)}{\int_{V_{\rm acc}} d{\bf r}\,e^{-\beta u(r-L)}}.
\label{eq8}
\end{equation}

\noindent
Here $\beta=1/k_BT$, $V_{\rm acc}$ is the volume accessible to the pseudo neighbor, and $y(r)$ is a step function that takes into account that the potential is cut off at 2.5$\sigma_{\alpha \beta}$, i.e.~$y(r)=1$ if $L\leq r \leq L+2.5\sigma_{\alpha\beta}$ and $y(r)=0$ for all other values of $r$. The volume integrals in Eq.~(\ref{eq8}) can be decomposed into a spherical part that is contained inside the cubic box, and the rest. The latter volume is given by

\begin{figure}[tb]
\includegraphics[width=0.45\textwidth,trim = {0 0cm 0 0.1cm},clip]{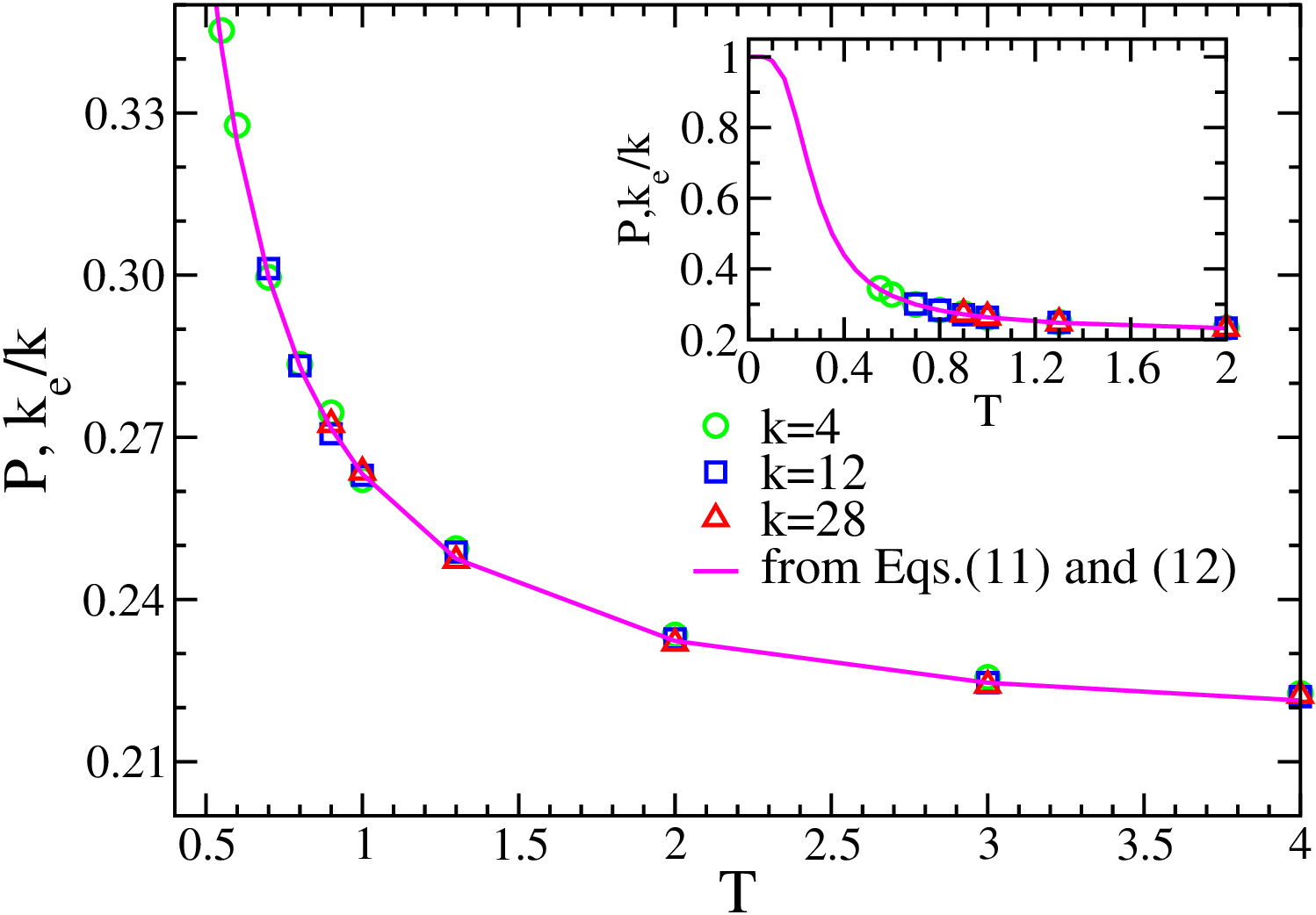}
\caption{Probability that a pseudo neighbour is within the interaction range as a function of temperature. The pink line is the theoretical prediction from Eqs.~(\ref{eq12}) and (\ref{eq13}). Inset: Same quantities extending the temperature range to $T=0$. The theoretical curve shows a sigmoidal shape.
}
\label{fig3_pseudo_5}
\end{figure}

\begin{eqnarray}
\Delta V&=&L_{\rm box}^{3}-\frac{4}{3}\pi \Big(\frac{L_{\rm box}}{2}\Big)^{3}\\
&=&L_{\rm box}^{3}(1-\frac{\pi}{6}) \quad .
\label{eq9}
\end{eqnarray}

A spherical integration in Eq.~(\ref{eq8}) gives then

\begin{equation}
P_{L}=\frac{\int_{L}^{L+r_c}dr\;r^{2}e^{-\beta u(r-L)}}{\int_{L}^{L_{\rm box}/2}dr\;r^{2}e^{-\beta u(r-L)}+\Delta V} \quad .
\label{eq10}
\end{equation}

Note that in the above expression, $L=L_{ij}$ is fixed. Hence for a  distribution of $L$, the probability of finding a pseudo neighbour within the interaction range of the tagged particle is given by

\begin{equation}
P=\int_{r_c}^{L_{\rm max}} \!\!\! dL\mathscr{P}(L)\frac{\int_{L}^{L+r_c}dr\; r^{2}e^{-\beta u(r-L)}}{\int_{L}^{L_{\rm box}/2} dr\;r^{2}e^{-\beta u(r-L)}+\Delta V}.
\label{eq11}
\end{equation}

In the numerator we make the substitution $r'=r-L$ which allows to interchange the two integrals:

\begin{equation}
\begin{aligned}
P=\int_{0}^{r_c}\!\! dr'\int_{r_c}^{L_{\rm max}}\!\!dL\mathscr{P}(L)\frac{(r'+L)^{2}e^{-\beta u(r')} }{\int_{L}^{L_{\rm box}/2}dr\,r^{2}e^{-\beta u(r-L)}+\Delta V}.
\end{aligned}
\label{eq12}
\end{equation}

\begin{figure}[tb]
\includegraphics[width=0.45\textwidth,trim = {0 0cm 0 0.1cm},clip]{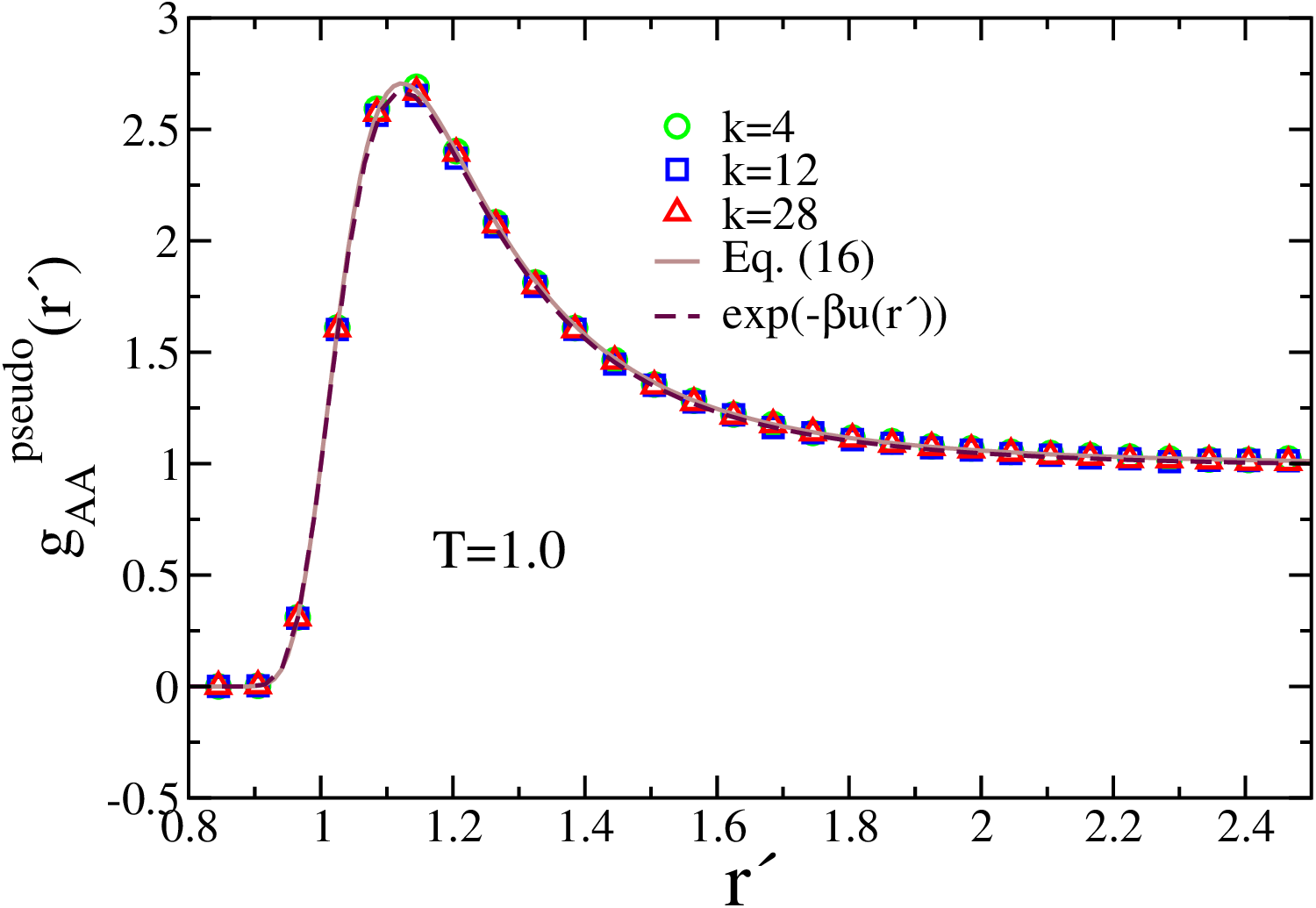}
\caption{Radial distribution function for pseudo neighbours from simulations at $T=1.0$ for $k=4,12$ and 28. The distribution function of the pseudo neighbours is independent of $k$. The solid line is the result from the theoretical expression given by Eq.~(\ref{eq17}). The dashed line is the theoretical prediction from the bare potential.
}
\label{fig4_gr_extra_compare_simu_all_k}
\end{figure}

We thus find that this probability is independent of $k$, a result that is reasonable since we have neglected any correlations between the pseudo neighbors.
Also note that $P$ depends on the interaction potential via $u(r)$ and $r_c$. For a binary system, we can generalize this calculation to obtain the partial probabilities $P_{\alpha\beta}$ and then the total probability is given by

\begin{equation}
P=x_{A}^2P_{AA}+2x_{A}x_{B}P_{AB}+x_{B}^2P_{BB}\quad ,
\label{eq13}
\end{equation}

\noindent
where $x_\alpha$ is the concentration of species $\alpha$. In the simulation, this probability can be obtained by calculating the ratio $k_e/k$, where $k_e$ is the number of pseudo neighbors that have a non-zero interaction with the tagged particle. In Fig.~\ref{fig3_pseudo_5} we show the temperature dependence of $P$ as obtained from Eqs.~(\ref{eq12}) and (\ref{eq13}) (solid line) and compare it with the corresponding quantity $k_e/k$ determined from the simulations (symbols).
One recognizes that $k_e/k$ is as expected independent of $k$ and that the simulation data matches perfectly well the theoretical prediction given by Eqs.~(\ref{eq12}) and (\ref{eq13}). Note that at the lowest temperatures at which we could equilibrate the systems for the different value of $k$ the probability is around 0.3, i.e.~for the glassy dynamics we will discuss below only a relatively  small part of the pseudo neighbors  are actually interacting with the tagged particle. The inset of the figure shows that $P$ becomes 0.5 at around $T=0.4$, a temperature at which already the $k=0$ system is very
viscous~\cite{coslovich_ozawa_kob}, and for $T\rightarrow 0$ the probability becomes 1, as expected.

To characterize the relative position of a pseudo-neighbor $j$ with respect to a tagged particle $i$ we can consider the corresponding radial distribution function

\begin{equation}
g^{\rm pseudo}(r') = \frac{\rho_k}{4\pi r^2}
\sum_{i=1}^N\sum_{j(i)}^k \langle \delta(r'-|{\bf r}_i-{\bf r}_j|+L_{ij})\rangle \quad,
\label{eq14}
\end{equation}

\noindent
where in the second sum the index runs over the pseudo neighbors of the tagged particle $i$ and $\rho_k$ is the average pseudo neighbour density, 

\begin{equation}
\rho_{k}=\int_{r_c}^{L_{\rm max}}\frac{k \mathscr{P}(L)}{V-\frac{4}{3}\pi L^{3}}dL \quad,
\label{eq15}
\end{equation}

\noindent
where $V$ is the total volume of the system.

To calculate $g^{\rm pseudo}(r)$ analytically we can make use of our result for $P$ given by Eqs.~(\ref{eq12}) and (\ref{eq13}). The number $k_e$ of pseudo neighbours within the interaction range can be expressed in terms of $g^{\rm pseudo}(r')$ as

\begin{equation}
k_{e}=\rho_{k} \int_{0}^{r_c} dr' g^{\rm pseudo}(r') \int_{r_c}^{L_{\rm max}}dL\mathscr{P}(L) 4\pi (r'+L)^{2}.
\label{eq16}
\end{equation}

Since $k_e$ can also be written as $k_{e}=k\times P$ we get, using Eq.~(\ref{eq12}) and Eq.~(\ref{eq16})

\begin{equation}
\begin{aligned}
&g^{\rm pseudo}(r')\rho_{k}\int_{r_c}^{L_{\rm max}}dL\mathscr{P}(L) 4\pi (r'+L)^{2} \\
&=k\int_{r_c}^{L_{\rm max}}dL\mathscr{P}(L)\frac{(r'+L)^{2}e^{-\beta u(r')}}{\int_{L}^{L_{box}/2}drr^{2}e^{-\beta u(r-L)}+\Delta V}
\end{aligned}
\label{eq17}
\end{equation}

\noindent
from which one obtains directly $g^{\rm pseudo}(r')$. Note that $g^{\rm pseudo}(r')$ is independent of $k$, since $\rho_k$ is directly proportional to $k$, see Eq.~(\ref{eq15}).

Fig.~\ref{fig4_gr_extra_compare_simu_all_k} shows the radial distribution function $g^{\rm pseudo}(r')$ from the simulations of three different values of $k$ (symbols) and we recognize that, as predicted by Eq.~(\ref{eq17}) the function is indeed independent of $k$. We have also included the analytical result from Eq.~(\ref{eq17}) and we see that the theory describes perfectly well the simulation data, thus demonstrating that the approximation that the structure of the pseudo neighbors can be obtained well by the bare interaction with the tagged particle is very accurate, at least for the $k$ values considered in the present work. We also note that since one has the relation  $g^{\rm pseudo}(r')= \exp(-\beta u(r'))$, which can be derived from Eq.~(\ref{eq17}), the function $g^{\rm pseudo}(r')$ can also be obtained directly from the bare interaction potential $u(r')$ as shown in Fig.\ref{fig4_gr_extra_compare_simu_all_k}.

Within the standard theory of liquids, the radial distribution function allows to obtain the potential energy~\cite{hansen_mcdonald_86}. Due to the presence of the pseudo neighbors this is no longer possible, and thus the usual expression has to be modified as follows. (Note that in the following we give the expressions for a one-component system. For the binary system considered here, one will have to do the sum over the various partials.)
Since the potential energy of the system has two contributions, one is the regular neighbour and the other the pseudo neighbour (see Eq.~(\ref{eq1})), the total potential energy $U_{\rm tot}$ is given by,

\begin{equation}
\begin{aligned}
&\frac{U_{\rm tot}}{N}=\frac{\rho}{2} \int_0^\infty u(r)g(r)4\pi r^{2}dr \hspace{140pt}\\
&+\frac{\rho_{k} }{2}\int_0^\infty u(r)g^{\rm pseudo}(r)\int_{r_c}^{L_{\rm max}}\mathscr{P}(L)4\pi (r+L)^{2} dL dr .
\label{eq18}
\end{aligned}
\end{equation}

At this stage it is useful to introduce an ``effective radial distribution'' function $g^{\rm eff}(r)$ by defining

\begin{equation}
\rho_{\rm eff}g^{\rm eff}(r)=\rho g(r) 
+{\rho_{k} }g^{\rm pseudo}(r)\frac{\int_{r_c}^{L_{\rm max}}\mathscr{P}(L) (r+L)^{2}dL}{r^{2}},
\label{eq19}
\end{equation}

\noindent 
where the effective particle density is given by

\begin{equation}
\rho_{\rm eff}=\rho+\rho_k\quad .
\label{eq20}
\end{equation}

Note that since $\rho_k$ increases linearly with $k$, for large $k$ the density $\rho_{\rm eff}$ is dominated by $\rho_k$ and hence in that limit $g^{\rm eff}$ will be directly proportional to $g^{\rm pseudo}(r)$.

Using $g^{\rm eff}(r)$ we now can express the total potential energy of the system as a function of the radial distribution function $g^{\rm eff}(r)$:

\begin{equation} 
\frac{U_{\rm tot}}{N}=\frac{\rho_{\rm eff}}{2} \int_0^\infty u(r)g^{\rm eff}(r)4\pi r^{2}dr \quad . 
\label{eq21}
\end{equation}

\begin{figure}[!ht]
\includegraphics[width=0.45\textwidth]{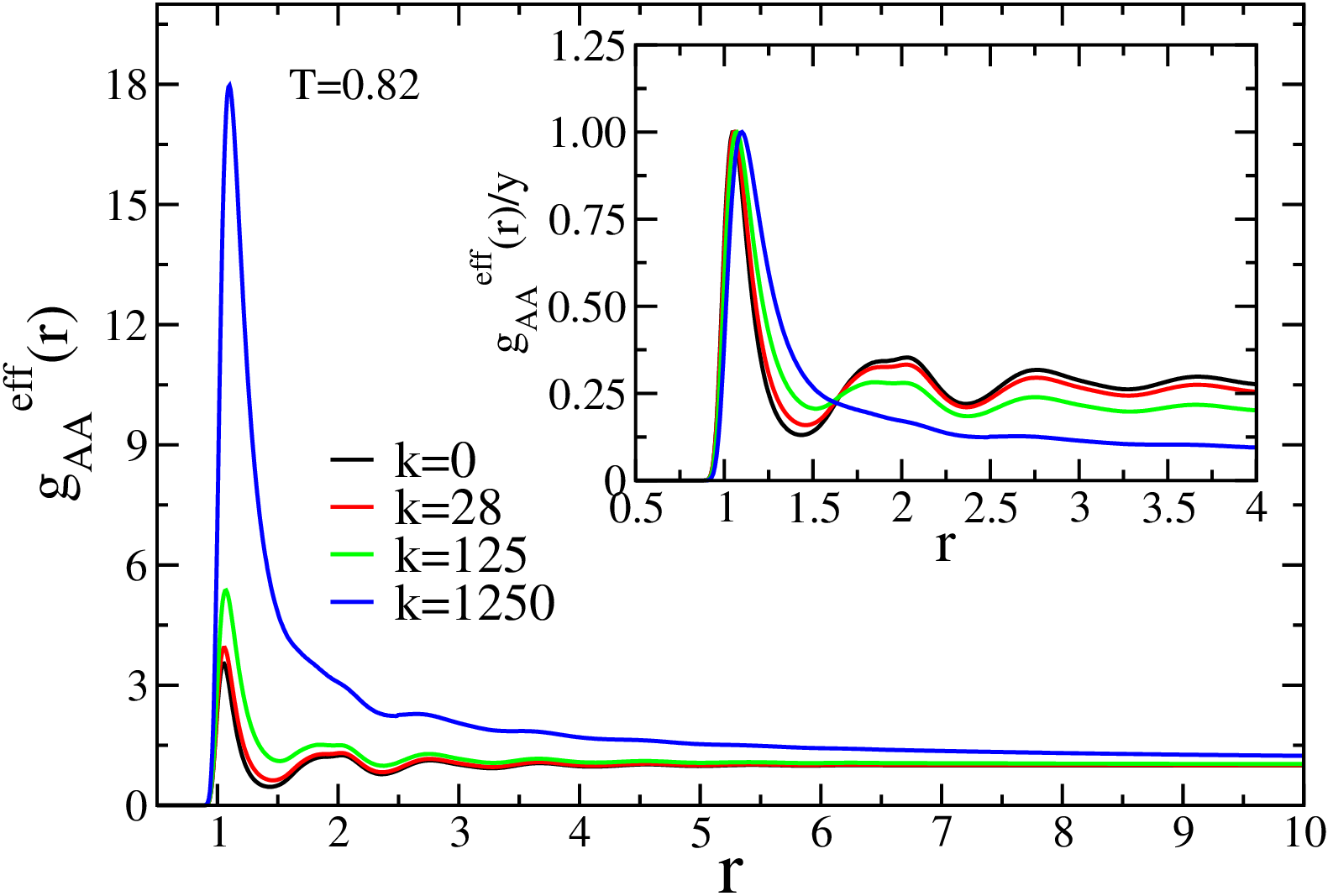}
\caption{The effective A-A particle radial distribution function for $k=0,28,125$, and 1250. With increasing $k$ the multi-peak structure disappears. Inset: $g^{\rm eff}_{AA}/y(k)$ vs $r$ where $y(k)$ is the height of the main peak. The smoothing of the undulation with increasing $k$ is clearly seen.
}
\label{fig5_gr_eff_fig}
\end{figure}

In Fig.~\ref{fig5_gr_eff_fig} we present  $g^{\rm eff}(r)$ for the A-A correlation for different values of $k$. Since the regular radial distribution function $g(r)$ is independent of $k$ (see Fig.~\ref{fig1_gr_regular}) and $g^{\rm pseudo}(r)$ can be calculated analytically from Eq.~(\ref{eq17}) it is possible to obtain $g^{\rm eff}$ for arbitrary values of $k$. The graph shows that with increasing $k$, the radial distribution function loses its characteristic structure with the multiple peaks and converges toward a distribution that has a single peak at $r=1$. This result can be understood directly from Eq.~(\ref{eq19}) since for large $k$ the first term on the right-hand side vanishes (if divided by $\rho_{\rm eff}$) while the second term is $g^{\rm pseudo}(r)$ multiplied by an $r-$dependent factor that is independent of $k$. So we see that in the large $k$ limit the effective radial distribution function develops a dominant sharp peak at a finite distance. With decreasing temperature, this peak increases since most of the pseudo neighbors will condensate at the optimal distance $L_{ij}$. It is this growing peak that signals the increasing number of constraints in the system which induce the slowing down of the relaxation dynamics. This loss of structure of the radial distribution function is a typical signature of mean-field-like systems, such as the hard-sphere system of Ref.~[\onlinecite{mari-kurchan}]. (However, unlike the results in the present study, in the hard-sphere system there is no peak at $r=1$.) 

\subsection{Relaxation dynamics}
We now analyze how the presence of the pseudo neighbours affects the relaxation dynamics. To characterize this dynamics we consider the self part of the overlap function $Q(t)$ and the mean squared displacement (MSD) of a tagged particle, $\Delta r^2(t)$. The former observable is defined as

\begin{equation}
Q(t) =\frac{1}{N}\sum_{i=1}^{N} \langle \omega (|{\bf{r}}_i(t)-{\bf{r}}_i(0)|)\rangle \quad ,
\label{eq22}
\end{equation}

\noindent
where the function $\omega(x)$ is 1 if $0\leq x\leq a$ and $\omega(x)=0$ otherwise. The parameter $a$ is chosen to be 0.3, a value that is slightly larger than the size of the cage (determined from the height of the plateau in the MSD at intermediate times~\cite{kob-andersen}.) Thus the quantity $Q(t)$ tells whether or not at time $t$ a tagged particle is still inside the cage it occupied at $t=0$.

\begin{figure}[!ht]
\includegraphics[width=0.45\textwidth,trim = {0 0cm 0 0.1cm},clip]{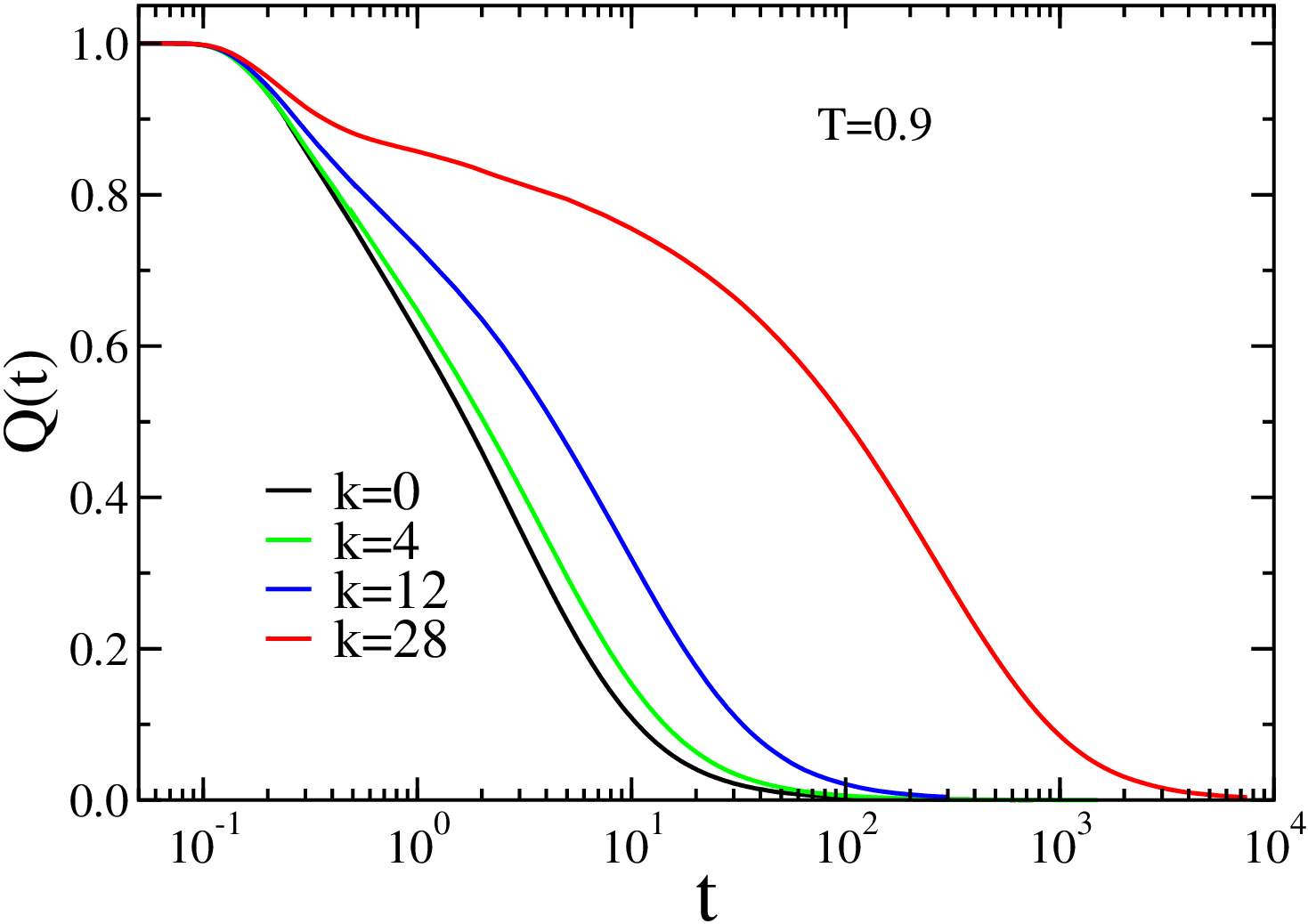}
\caption{Time dependence of the self part of overlap function $Q(t)$ for systems with different values of $k$ at $T=0.9$. With increasing $k$ the relaxation dynamics quickly slows down.
}
\label{fig6_overlap_diff_k}
\end{figure}

In Fig.~\ref{fig6_overlap_diff_k} we show the time dependence of $Q(t)$ for different values of $k$. The temperature is $T=0.9$ which corresponds for $k=0$ to a $T$ that is around the onset temperature~\cite{kob-andersen,atreyee_onset}. The graph demonstrates that with increasing $k$, the relaxation dynamics slows down quickly, in that the correlator for $k=28$ decays on a time scale that is about two orders of magnitude larger than the one for $k=0$.
Also note that for the largest $k$ we clearly see a two-step relaxation, i.e., the hallmark of glassy dynamics in which the particles are temporally trapped by their neighbors~\cite{binder_kob_book}, while for $k=0$ one has just a simple one-step relaxation, i.e., a normal liquid state relaxation.
These results demonstrate that the presence of the pseudo neighbors does  have the sought after effect of strongly slowing down the relaxation dynamics of the system, although, as demonstrated above, the overall structure of the liquid is not changed. Interestingly the shape of the time correlation function in the $\alpha$-relaxation regime does not seem to have a noticeable dependence on $k$, indicating that the relaxation mechanism is weakly dependent on $k$.
However, this conclusion only holds for length scales on the order of $'a'$ while it could be that on larger scales differences become noticeable. 
Here we also note that for other mean-field like models, such as the one introduced by Mari and Kurchan~\cite{mari-kurchan}, an increase of the interaction range leads to an {\it acceleration} of the dynamics, i.e.~the hoped for slowing down of the dynamics is not necessarily guaranteed.

\begin{figure}[!b]
\includegraphics[width=0.5\textwidth,trim = {0 0cm 0 0.1cm},clip]{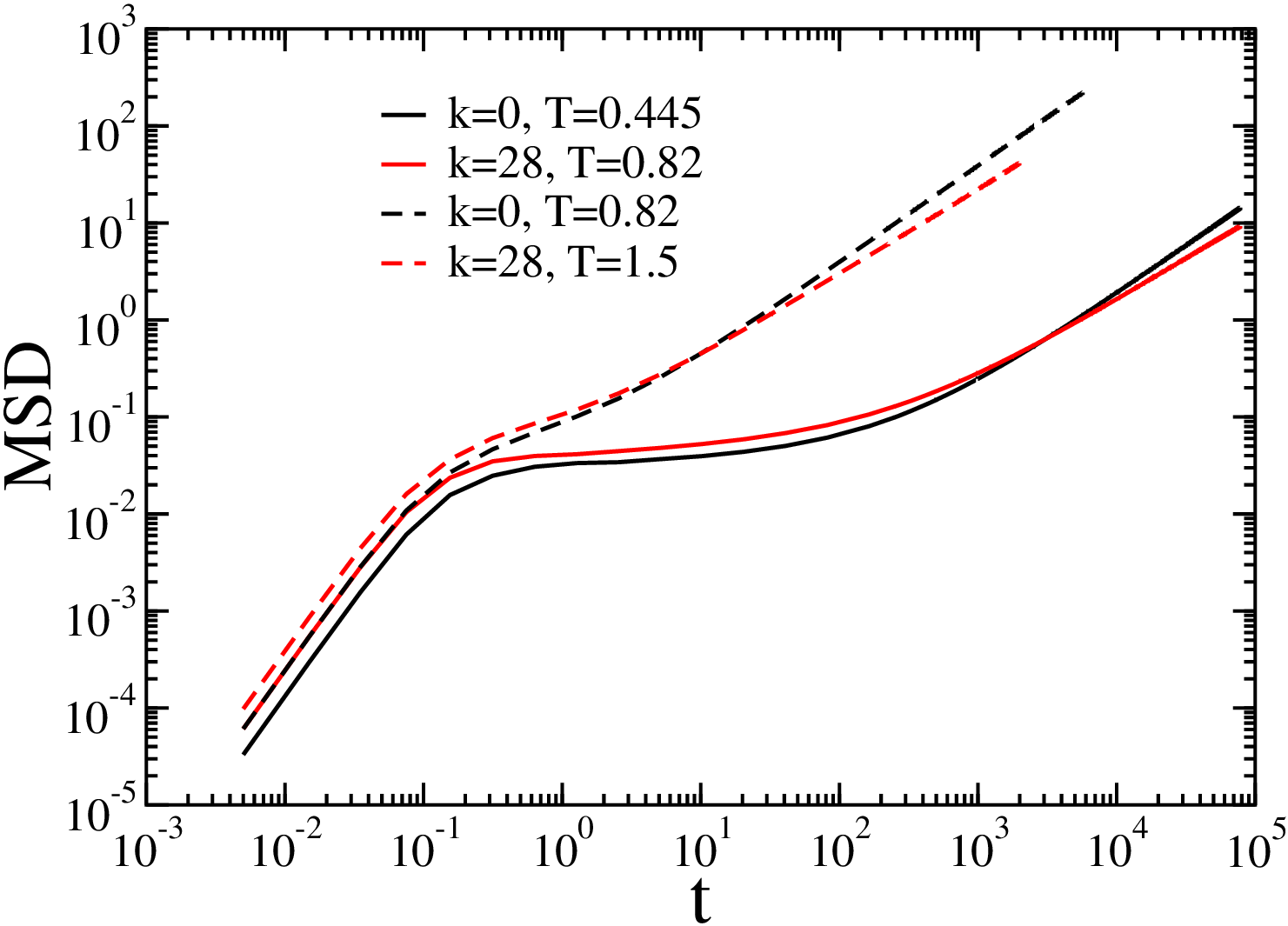}
\caption{Time dependence of the mean squared displacement for the $k=0$ and $k=28$ systems in the high and low temperature regimes. The curves are for similar value of relaxation time. The $k=28$ system shows a weak sub-diffusive behaviour at high and low temperature.
} 
\label{fig7_msd}
\end{figure}

\begin{figure*}[!ht]
\centering
\begin{subfigure}[h]{0.4\textwidth}
\hspace*{-93mm}
\includegraphics[width=0.9\textwidth,trim = {0 0cm 0 0.1cm},clip]{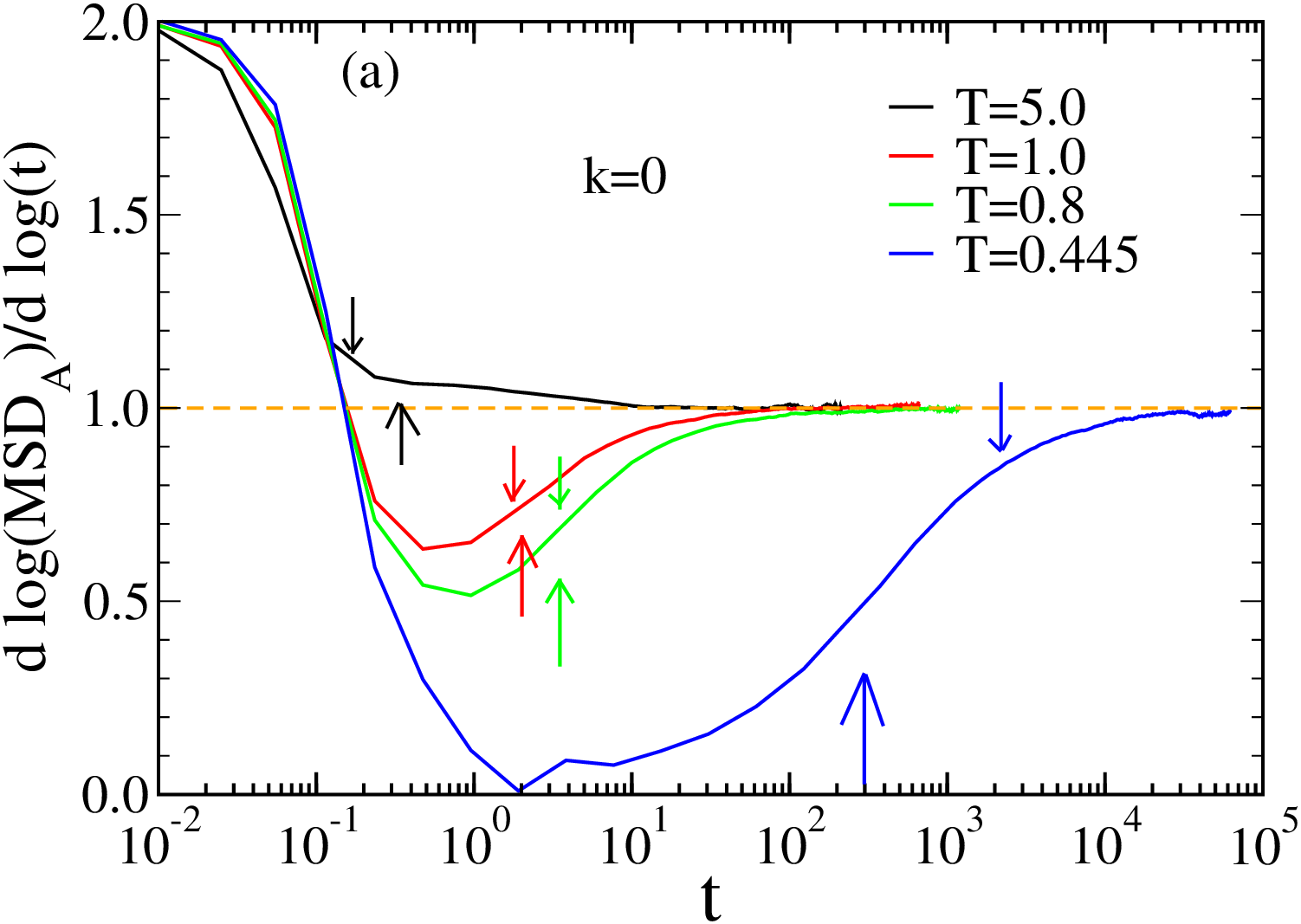}

\end{subfigure}

\begin{subfigure}[t]{0.4\textwidth}
\vspace*{-46mm}
\hspace*{47mm}
\includegraphics[width=0.9\textwidth,trim = {0 0cm 0 0.1cm},clip]{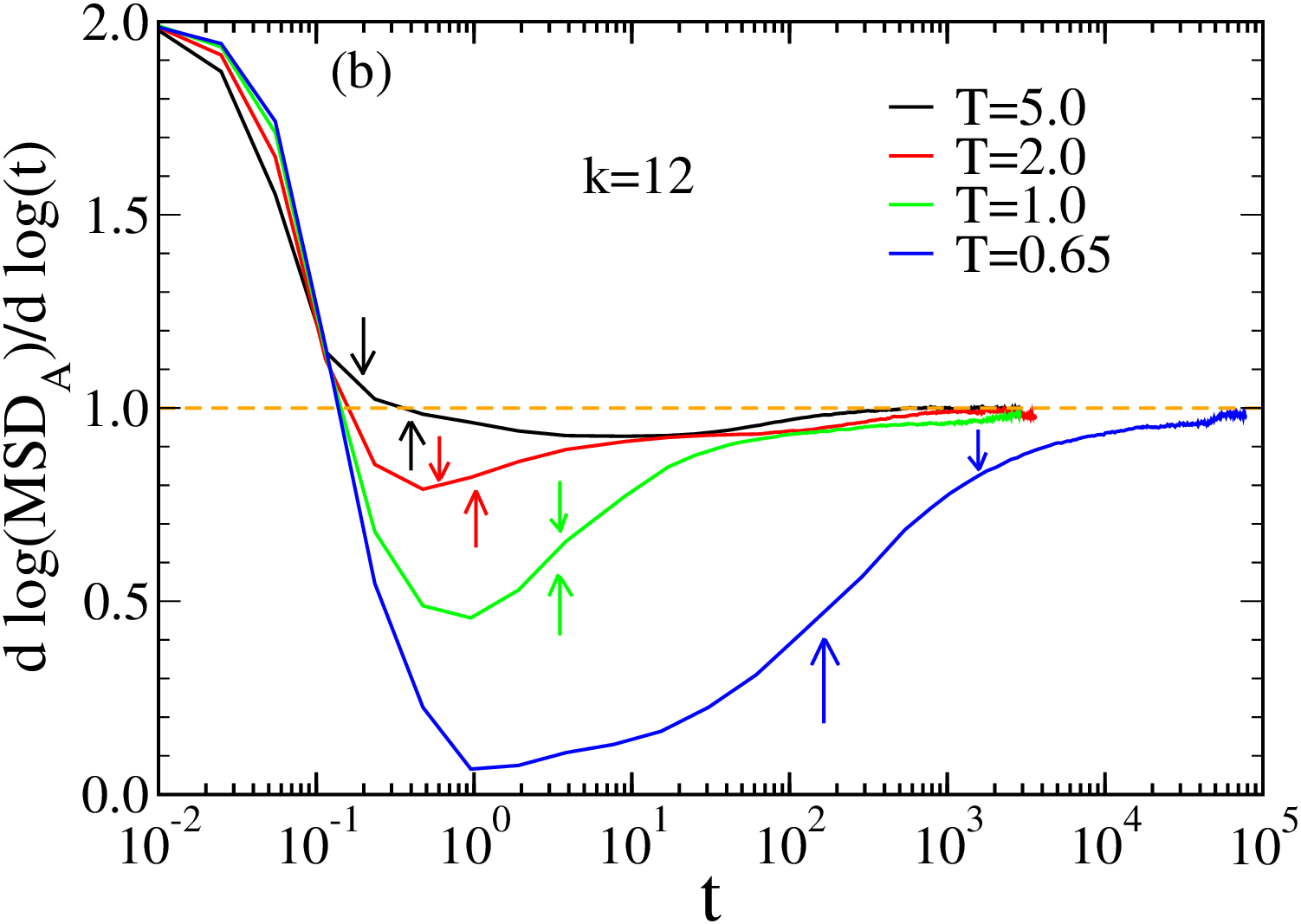}
\end{subfigure}
\begin{subfigure}[t]{0.6\textwidth}
\hspace*{-101mm}
\includegraphics[width=0.6\textwidth,trim = {0 0cm 0 0.1cm},clip]{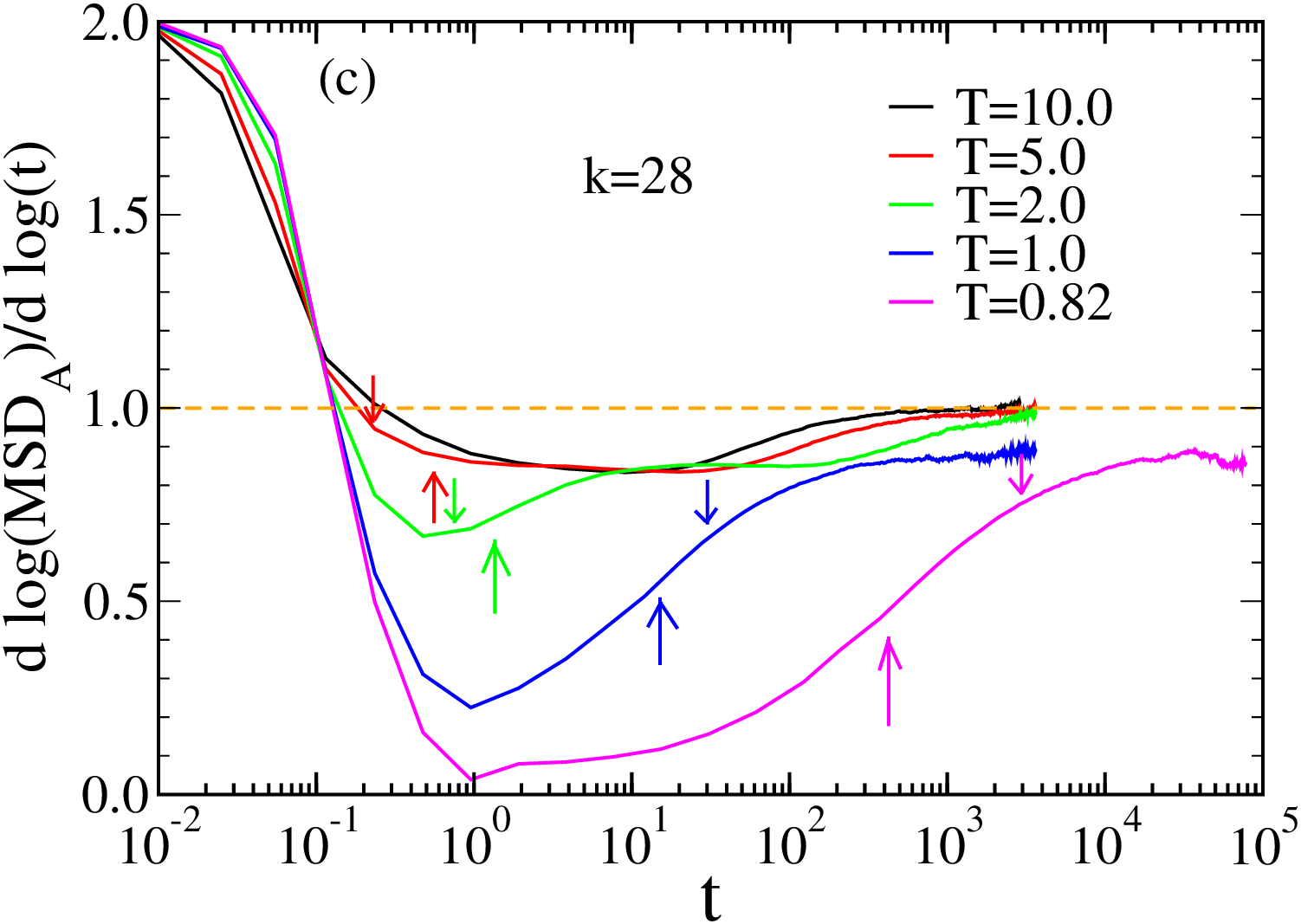}
\end{subfigure}
\hspace*{-100mm}
\begin{subfigure}[t]{0.5\textwidth}
\hspace*{50mm}
\includegraphics[width=0.73\textwidth,trim = {0 0cm 0 0.1cm},clip]{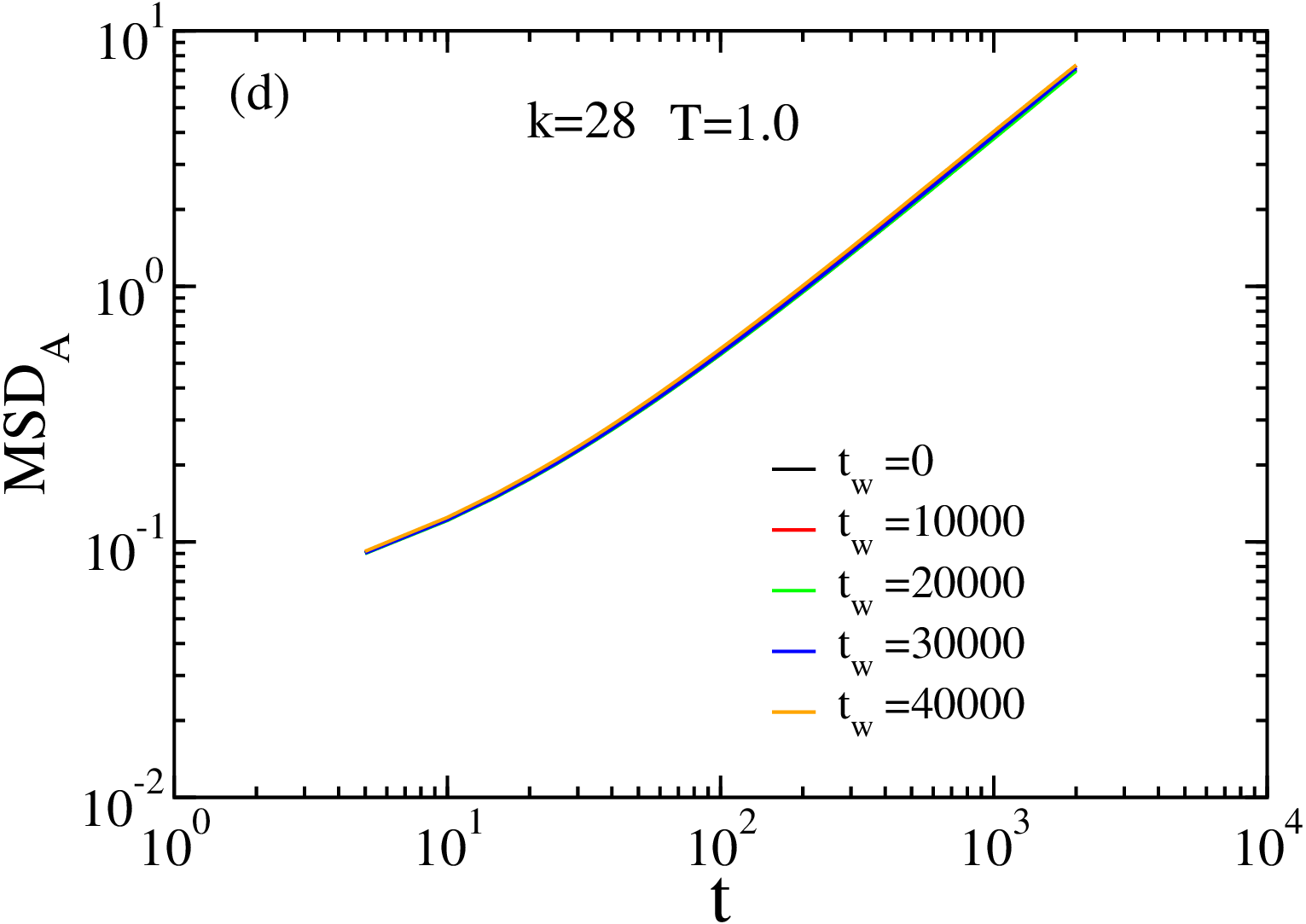}
\end{subfigure}
\caption{Double-logarithmic derivative of the MSD of the A particles as a function of time. (a) System for $k=0$. If temperature is decreased the derivative shows at low $T$ a local minimum, indicating the presence of caging. (b) System for $k=12$. Qualitatively the same time dependence as in panel (a) but now at higher temperatures. (c) System for $k=28$. One sees that the curves show at intermediate times a plateau that is due to the caging caused by the pseudo neighbors. 
The arrows pointing upward [downward] in panels (a)-(c) indicate $\tau_2$ [$\tau_4$], the location of the peak in the non-Gaussian parameter $\alpha_2(t)$ [in the dynamic susceptibility $\chi_4(t)$]. (d) MSD of the A particles for different waiting times $t_w$ (see legend). No waiting time dependence is noticeable.
}
\label{fig9_msd_slope}
\end{figure*}

Next, we compare the time dependence of the mean squared displacement, averaged over all the particles, of two systems, $k=0$ and $k=28$, Fig.~\ref{fig7_msd}. For the $k=0$ system we show the MSD for $T=0.82$, i.e., a temperature close to the onset $T$ and as a consequence one sees that the curve shows between the ballistic regime at short times, $\Delta r^2(t)\propto t^2$, and the diffusive regime at long times, $\Delta r^2(t) \propto t^1$, a weak shoulder. Qualitatively the same time-dependence is found for the $k=28$ system, but this time at the higher temperature, $T=1.5$, indicating that the increase of $k$ leads to an increase of the onset temperature. If for the $k=0$ system the temperature is lowered to 0.445, the MSD shows at intermediate times a very pronounced plateau that is due to the temporary caging of the particles~\cite{binder_kob_book}. The same behavior is found in the $k=28$ system at $T=0.82$ with a plateau height and length that is very close to the one of the $k=0$ system. (This similarity is due to our choice of the temperature $T=0.82$). Since we have seen above that the local structure of the system at fixed temperature hardly depends on $k$, see Fig.~\ref{fig1_gr_regular},  the pronounced caging for the $k=28$ system (at T=0.82) is thus due to the pseudo neighbors, i.e., the non-local interactions. From these curves we hence can conclude that the presence of the additional interactions leads to a substantial slowing down of the relaxation dynamics while the details of the MSD, such as the height of the plateau or its width, at the same effective temperature (discussed below) are modified only mildly, at least in the parameter regime probed here.

\begin{figure}[th]
\includegraphics[width=0.45\textwidth]{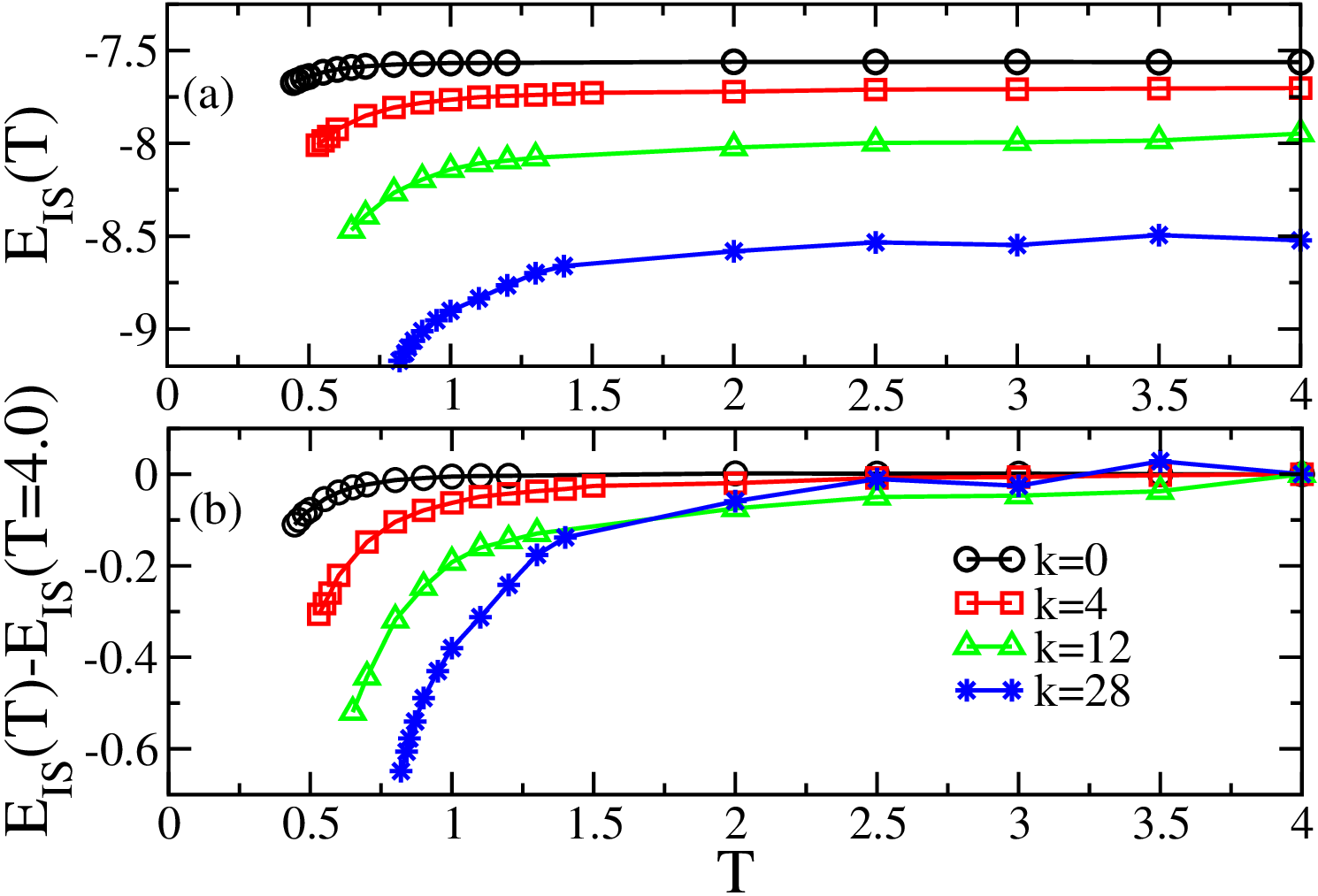}
\caption{(a) Inherent structure energy, $E_{\rm IS}$, as a function of temperature for the $k=0,4,12,$ and 28 systems.  (b) Shifted (by $E_{\rm IS}(T=4.0)$) inherent structure energy vs. $T$. Near $T_{\rm onset}$ the energy starts to deviate from its high temperature value allowing to determine $T_{\rm onset}$. With increasing $k$, $T_{\rm onset}$ moves to higher temperatures.\\
}
\label{fig10_EIS}
\end{figure}

At sufficiently long times the motion of the particles is expected to be diffusive, and hence the MSD should increase linearly in time. Fig.~\ref{fig7_msd}
shows that for the $k=0$ system, this is indeed the case and that this diffusion sets in once the MSD has reached a value around 1.0. Interestingly one observes for the $k=28$ system  even at the longest times a sub-diffusive behavior, with an exponent that is around 0.8, and this even for values of the MSD that are on the order of 10. This behavior can be noticed better by calculating the slope of the MSD in the log-log presentation, see Fig.~\ref{fig9_msd_slope}. For $k=0$, panel (a), we see that at short times the slope is 2.0, as expected for a ballistic motion. At high temperatures the slope crosses over to 1.0 at around $t=3$, i.e.~the system becomes diffusive. If $T$ is lowered, the slope starts to show a dip with a depth and width that increase rapidly with decreasing temperature. For long times we see, however, that the curves again attain the value of 1.0, i.e. the system is diffusive. Qualitatively the same behavior is found for $k=4$ (not shown) and $k=12$, panel (b). 
However, a closer inspection of the curve for $T=2.0$ reveals that after the first dip in the slope, the curve does not rise immediately to the value 1.0 but shows instead a plateau at a height of around 0.9 in the time window $5\leq t \leq 200$. The asymptotic value 1.0 is thus reached only at longer times, i.e.~the MSD shows a sub-diffusive regime. Qualitatively the same behavior is found for $k=28$, panel (c), but now the mentioned plateau at intermediate times becomes more visible since its height has decreased to 0.8, i.e. the deviation from the diffusive regime become more pronounced. We now clearly see that if the temperature is lowered the curves reach this second plateau at a later time, but its height is unchanged (see the curves for $T=1.0$ and 0.82).
Note that this plateau at long times is indeed a distinct dynamic regime and not just a brief transient during which the system approaches the diffusive limit. We also exclude the possibility that this new plateau is just an out-of-equilibrium phenomenon since, see panel (d), the MSD for different waiting times show no waiting time dependence.
We interpret this new regime as a consequence of the interaction of the tagged particle with its pseudo neighbors. These interactions will vanish only if all the involved pairs have moved by a radial distance of around $r_c$, and, because of geometrical reasons 
(the volume of the spherical cap increases with $L_{ij}$) and the fact that $L_{ij}>r_c$, this takes certainly more time than cutting just the interactions between the tagged particle and its nearest neighbors, which explains the long time tail in the MSD.
Note, however, that for sufficiently long times the MSD can be expected to become diffusive for all values of $k$, see, e.g.,~the curve for $T=2.0$ in panel (c).
This behaviour is thus similar to that observed earlier in systems where there are two length-scales \cite{kob-pinaki-pre}. 
In order to distinguish in the following the two mentioned processes, we will refer to the one corresponding to the particles leaving their nearest neighbor cage as the ``NN-$\alpha$-process'', while the dynamics in which the pseudo-neighbors leave the interaction range of the tagged particle will be referred to as the ``PN-$\alpha$-process''. 
Note that although Fig.~\ref{fig9_msd_slope} clearly indicates that there are two processes, we will see in the following that not all observables reveal this in a direct manner. For example, the time dependence of $Q(t)$, presented in Fig.~\ref{fig6_overlap_diff_k}, does not indicate an obvious presence of two different $\alpha-$processes, although the pseudo-neighbors can be expected to affect not only the relaxation time but also the details of the correlator.

\begin{figure}[th]
\centering
\begin{subfigure}[h]{0.38\textwidth}
\includegraphics[width=0.9\textwidth]{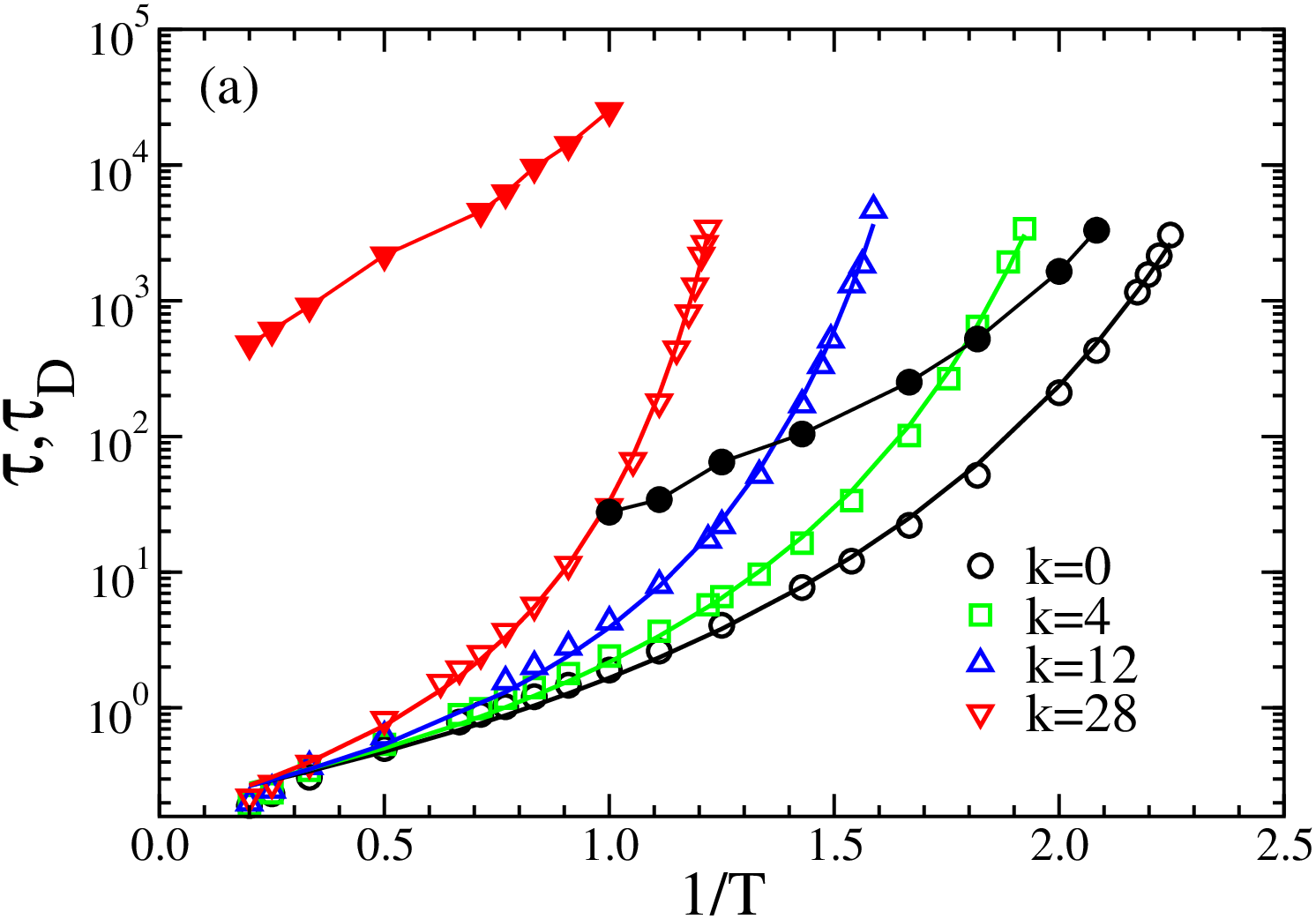}
\end{subfigure}
\begin{subfigure}[h]{0.38\textwidth}
\includegraphics[width=0.9\textwidth,trim = {0 0cm 0 0.1cm},clip]{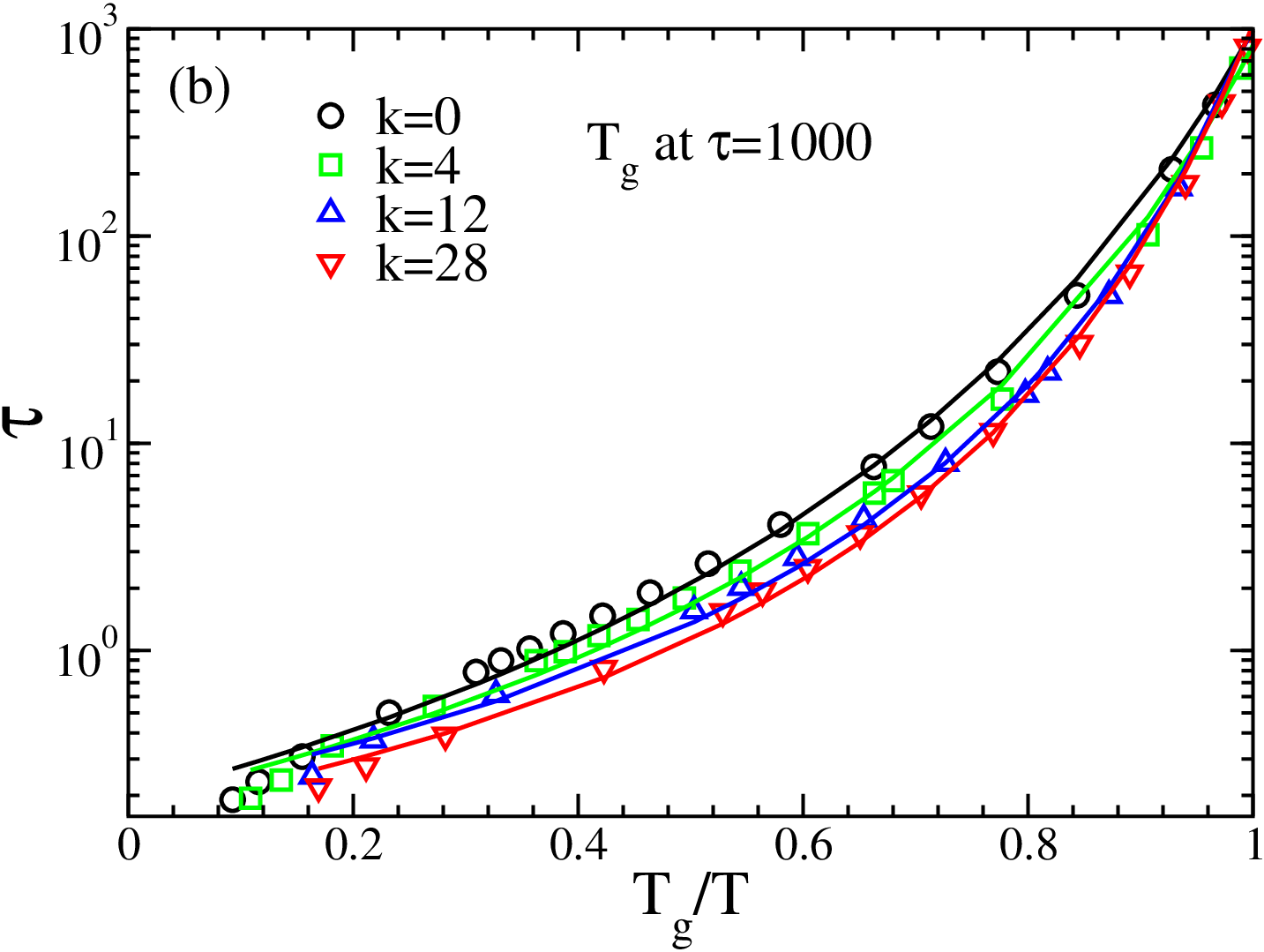}
\end{subfigure}
\begin{subfigure}[h]{0.38\textwidth}
\includegraphics[width=0.9\textwidth,trim = {0 0cm 0 0.1cm},clip]{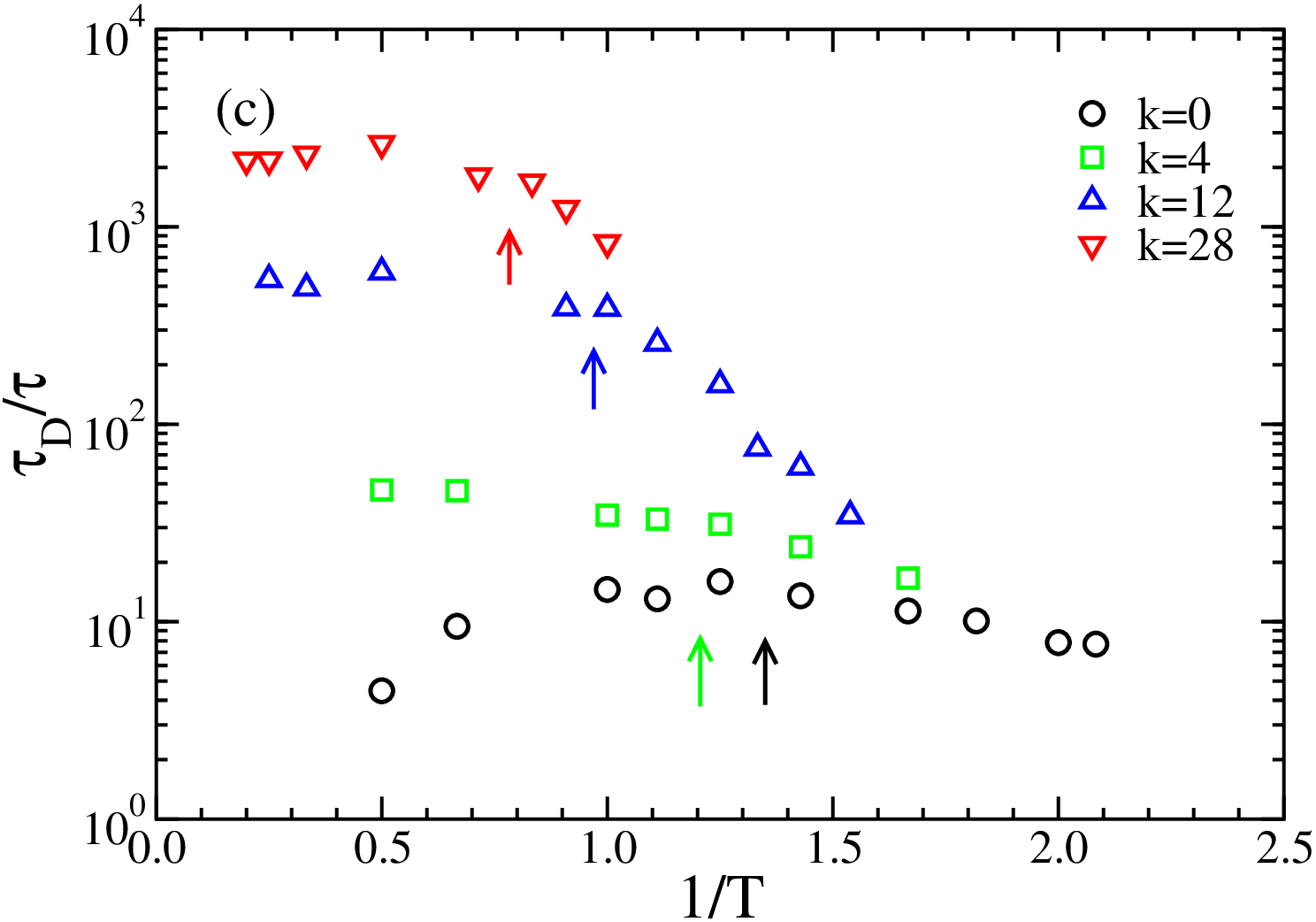}
\end{subfigure}
\caption{(a) Arrhenius plot of the $\alpha$-relaxation time, $\tau$, 
and the relaxation time obtained from the MSD, $\tau_{D}$, for systems with different values of $k$. Open and full symbols are for $\tau$ and $\tau_D$, respectively. The lines are fits to $\tau$ with the Vogel-Fulcher-Tammann expression, Eq.~(\ref{eq24}). (b) Same data as in (a) but now as a function of the scaled temperature $T_g/T$, with $\tau(T_g)=10^3$.
(c) Temperature dependence of the ratio $\tau_{D}/\tau$ for different values of $k$. The arrows indicate $T_{\rm onset}$. }

\label{fig_vft_all_k}
\end{figure}

Since the onset temperature is an important point on the energy scale of the system, we now have a closer look at the $k$-dependence of $T_{\rm onset}$. As mentioned above, this temperature can be identified from the first occurrence of a plateau in the MSD. Alternatively one can study the inherent structure energy, $E_{\rm IS}$, which shows at $T_{\rm onset}$ a marked change in its $T$-dependence~\cite{sastry-nature,sastry-onset}. (We recall that $E_{\rm IS}$ of a configuration is the potential energy evaluated at the local minimum of the energy reached from the configuration via the steepest descent procedure.)
In Fig.~\ref{fig10_EIS}(a) we show $E_{\rm IS}$ as a function of $T$, with the different curves corresponding to different values of $k$. From the graph, one recognizes that with increasing $k$ the energy decreases, an effect that is due to the presence of the pseudo neighbors which can lower the energy by occupying the well in the interaction potential. Less trivial is the fact that the temperature at which the curve starts to decrease rapidly, i.e.~the onset temperature, increases with increasing $k$. Thus the increase of $T_{\rm onset}$ with $k$ can be seen directly from this static observable. In order to see better the $k$-dependence of $T_{\rm onset}$, we plot in Fig.~\ref{fig10_EIS}(b)
the inherent structure energy shifted by $E_{\rm IS}(T=4.0)$. (The choice of $T=4.0$ for this normalization is not crucial.) The resulting graph clearly shows that the bend in the inherent structure energy occurs at higher temperatures with growing $k$, demonstrating the increase of the onset temperature. Fitting two straight lines to the data for $T>T_{\rm onset}$ and $T<T_{\rm onset}$, their intersection point can be used to determine $T_{\rm onset}$. As we will show elsewhere~\cite{nandi_thermo}, the so obtained values are compatible with the values of onset temperature as determined from the entropy~\cite{atreyee_onset}.
In Table~\ref{table_compare_temp} we list the values of $T_{\rm onset}$ obtained from these curves and one sees that for $k=28$ this temperature is about 90\% higher than $T_{\rm onset}$ for $k=0$.

\begin{table*}
\caption{The value of the characteristic temperatures and the kinetic fragility parameter for systems with different values of $k$. $T_{\rm onset}$ is the onset temperature at which the inherent structure energy starts to deviate significantly from its high temperature value. $T_{c}$ is the MCT transition temperature. $T_0$ is the singular temperature of the Vogel-Fulcher-Tammann equation, Eq.(\ref{eq24}).
All characteristic temperatures increase with increasing $k$. Also included are the normalized differences between various temperatures.
$K$ is the kinetic fragility defined in Eq.~(\ref{eq24}). $x(k)$ is the prefactor needed for the scaling plot shown in Fig.~\ref{fig11_mct_fitting}(b).
}
\begin{center}
\renewcommand{\arraystretch}{1.5}
\begin{tabular}{ | c | c | c | c | c | c | c | c | c | c| p{1.5cm} |}
\hline
$k$ & $T_{\rm onset}$ & $T_c$ & $T_0$ & $\frac{T_{\rm onset}-T_c}{T_c}$ &  $\frac{T_{\rm onset}-T_c}{T_{\rm onset}}$  & $\frac{T_{\rm onset}-T_0}{T_0}$ & $\frac{T_c-T_0}{T_0}$ & $K$ & $x(k)$ \\ \hline
0 & 0.74 $\pm 0.04$ & 0.43 & 0.283 & 0.72 & 0.42 & 1.61 & 0.52  & 0.184 & 1.0 \\ \hline
4 & $0.83 \pm 0.08$  & 0.51 & 0.362 & 0.63 & 0.38 & 1.29 & 0.41  & 0.237& 1.55\\ \hline
12 & $1.03\pm 0.07$ & 0.62 & 0.465 & 0.66 & 0.40 & 1.22 & 0.33   & 0.286 & 2.0 \\ \hline
28 & $1.28\pm 0.22$ & 0.80  & 0.610 & 0.60 & 0.38 & 1.10 & 0.31  & 0.297 & 2.1 \\ \hline
\end{tabular}
\label{table_compare_temp}
\end{center}
\end{table*}

A further important quantity to characterize the relaxation dynamics of a glass-former is the $\alpha$-relaxation time $\tau$. Here we define this time scale via $Q(\tau) = 1/e$. This definition is reasonable since we have seen in Fig.~\ref{fig6_overlap_diff_k} that the shape of the time correlation functions is basically independent of $k$. 
(Note that with this definition of $\tau$ we do not distinguish between the NN-$\alpha$-process and the PN-$\alpha$-process discussed in the context of Fig.~\ref{fig6_overlap_diff_k}. For the values of $k$ considered here, this is justified since the final decay of $Q(t)$ involves both processes.)
Fig.~\ref{fig_vft_all_k}(a) is an Arrhenius plot of $\tau$ for the different systems. One clearly sees that with increasing $k$, the dynamics quickly slows down and that the bending of the curve seems to increase, i.e.~the system becomes more fragile. To quantify this trend as a function of $k$, we have fitted $\tau(T,k)$ at intermediate and low temperatures to a Vogel-Fulcher-Tammann(VFT)-law:

\begin{equation}
\tau(T) = \tau_{0}\exp\Big[\frac{1}{K(T/T_0-1)}\Big] \quad .
\label{eq24}
\end{equation}

Here $T_0$ is the so-called VFT temperature at which the relaxation time of the system is predicted to diverge. The parameter $K$ describes the curvature of the data in an Arrhenius plot and hence can be considered as a measure for the fragility of the glass-former. The figure demonstrates that this functional form gives a good fit to the data (solid lines) and hence allows to estimate $T_0$ and $K$.

The values of $T_0$ are included in Tab.~\ref{table_compare_temp} as well and one sees that $T_0$ changes by about a factor of two if $k$ is increased from 0 to 28, i.e.~a factor that is comparable to the one found for $T_{\rm onset}$. In contrast to this we find that the parameter $K$ occurring in the Vogel-Fulcher-Tammann-law, Eq.~(\ref{eq24}), increases by about 30\% in the considered $k$-range, see Tab.~\ref{table_compare_temp}. This indicates that the introduction of the pseudo neighbors renders the system increasingly more fragile. Another way to see this is to define an effective glass transition temperature $T_g$ via $\tau(T_g)=10^3$ and to plot the relaxation time as a function of $T_g/T$~\cite{binder_kob_book,angell}. This is done in Fig.~\ref{fig_vft_all_k}(b) and one sees that the curves for large $k$ are indeed more bent than the ones for small $k$, i.e.~the fragility of the system increases with $k$.
This trend is thus qualitatively similar to the observation of Ref.~\onlinecite{sastry-SE} in which it was found that increasing the dimensionality of a glass-former gives rise to a higher fragility. 

Since the MSD has shown that the system has two kind of $\alpha-$processes it is useful to study how the corresponding relaxation times relate to each other. For the $k>0$ systems particles are caged by their nearest neighbours  as well as by their pseudo neighbours. When a particle leaves its NN cage the overlap function decays and this timescale is captured by $\tau$. We now define a relaxation time $\tau_{D}$ for the PN-process as the time scale at which the system becomes diffusive, i.e the time where the logarithmic derivative of the MSD goes to 1 \cite{smarajit_tauD}. In practice we consider $t=\tau_{D}$ for which $\frac {{\rm d log(MSD)}}{{\rm d log} (t)}= 0.97$. In Fig.~\ref{fig_vft_all_k}(a)
we have included the $T$-dependence of $\tau_D$ for the $k=0$ and the $k=28$ systems and one recognizes that $\tau_D$ is significantly larger than $\tau$ but that its $T$-dependence is weaker. To see the latter in a clearer way we show in panel (c) the $T$-dependence of the ratio $\tau_D/\tau$ for all value of $k$ considered. 
We recognize that 
the ratio starts to decrease quickly for temperatures that are below $T_{\rm onset}$, i.e.~once the systems start to show glassy dynamics. Since this decrease is very pronounced for $k>0$, we conclude that the slowing down of the overall dynamics of the system is mainly governed by the NN~$\alpha$-process (which is strongly influenced by the presence of the pseudo neighbors). 

These results show that the pseudo neighbors strongly influence the relaxation dynamics of a tagged particle in that the leaving of the cage formed by the nearest neighbors is strongly slowed down, as indicated by $\tau(T)$. In addition the pseudo neighbors also induce a new slow process, the PN-$\alpha$ process, which is related to the motion of the pseudo neighbors with respect to the tagged particle. However, this slow process does not depend very strongly on $T$ since there is no structural correlation between the pseudo neighbors of a given tagged particle (this in contrast to the nearest neighbors which are correlated because of the local steric hindrance). As a consequence this slow PN-$\alpha$ process is {\it not} the mechanism responsible for the slowing down of the overall dynamics of the system. The relevant mechanism for this is thus given by the NN-$\alpha$ process.

\subsection{MCT power law}
Having presented our findings regarding the relaxation dynamics of the system we now probe whether this dynamics can be described by means of mode coupling theory. MCT predicts that close to the critical temperature $T_c$ of the theory the relaxation times show a power law divergence:

\begin{equation}
\tau(T)=\tau_{\rm MCT} (T-T_c)^{-\gamma} \qquad .
\label{eq23}
\end{equation}

\begin{figure}[!bth]
\begin{subfigure}[h]{0.4\textwidth}
\includegraphics[width=0.9\textwidth]{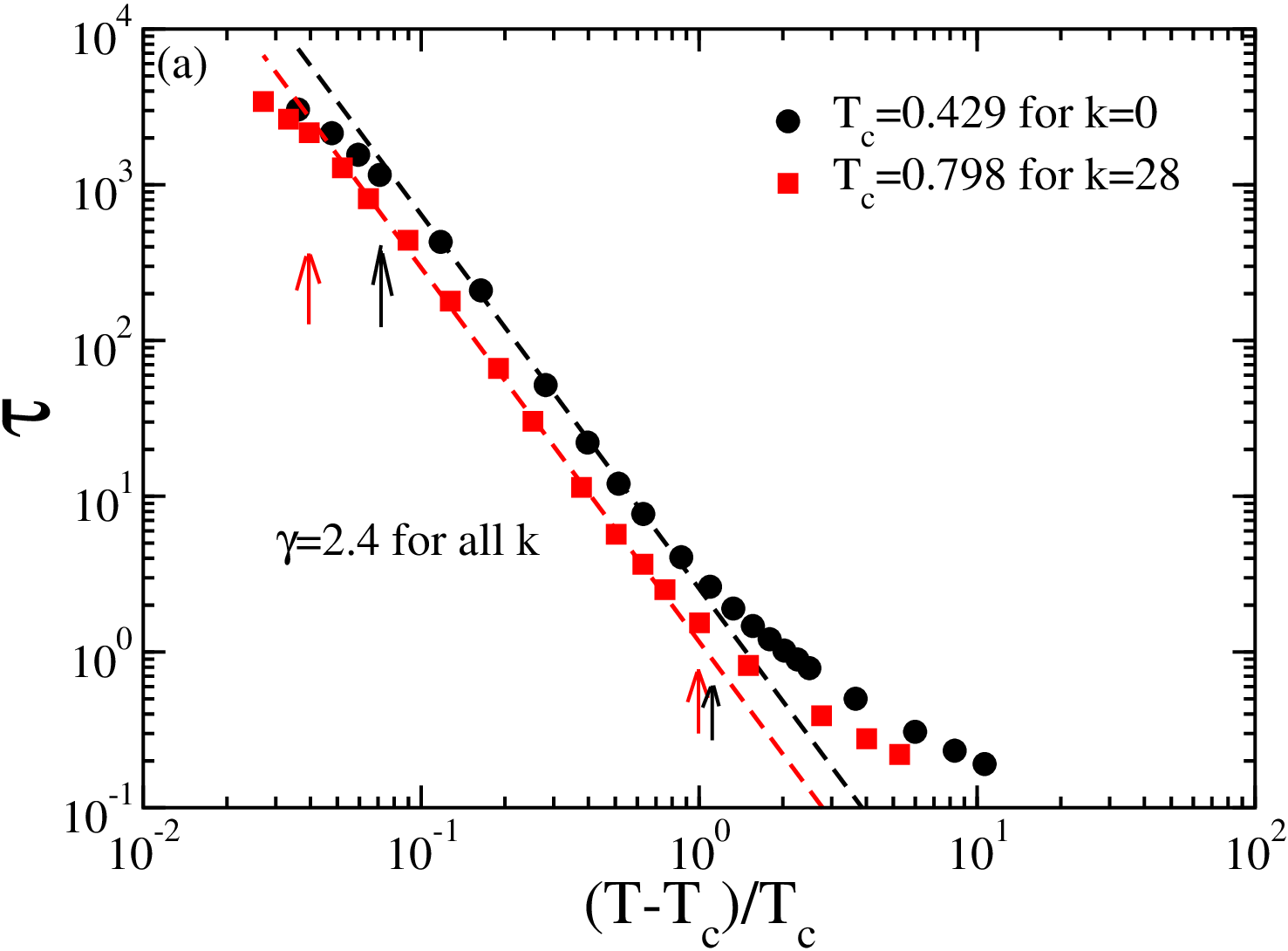}
\end{subfigure}
\begin{subfigure}[h]{0.4\textwidth}
\includegraphics[width=0.9\textwidth,trim = {0 0cm 0 0.1cm},clip]{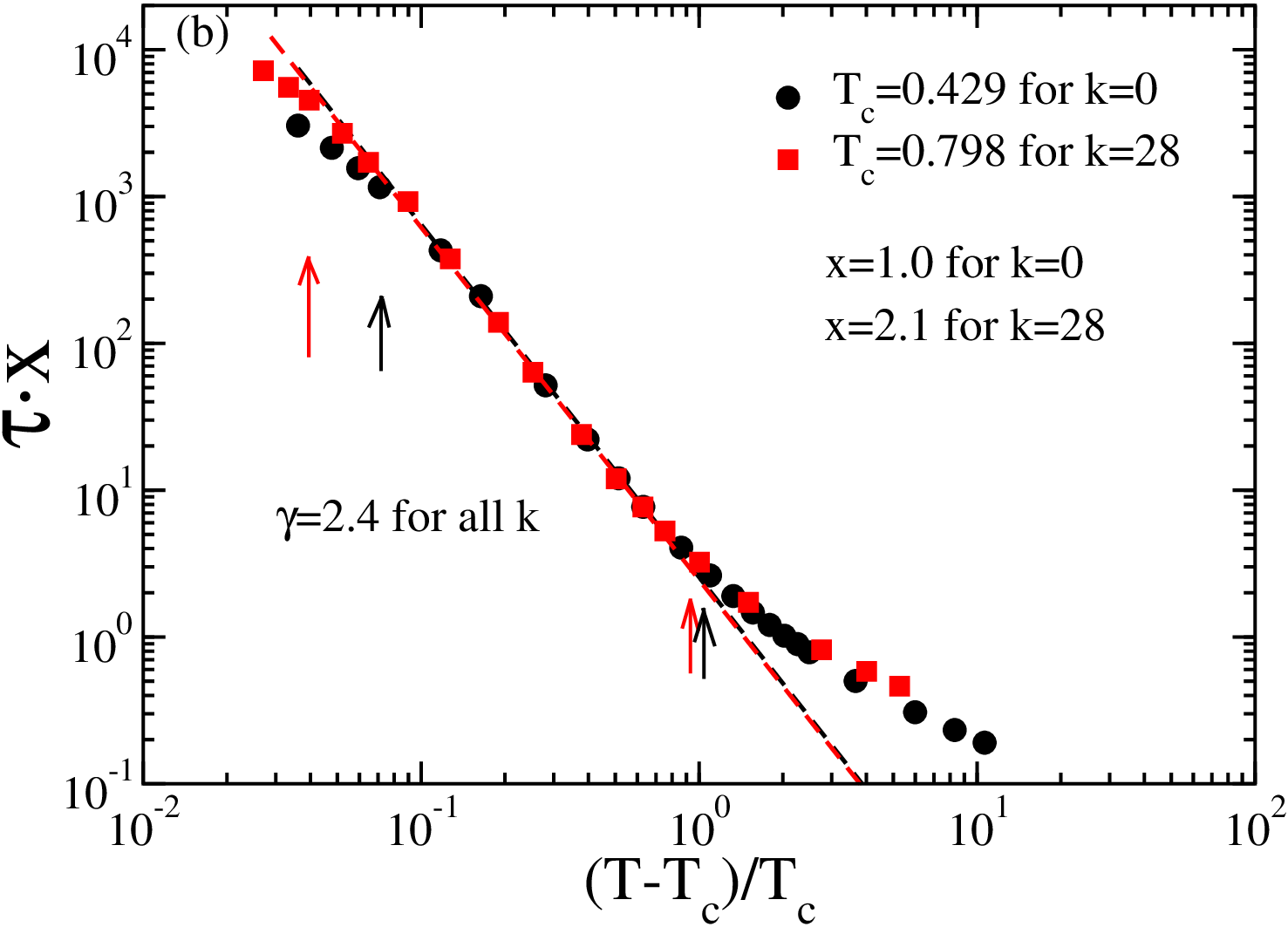}
\end{subfigure}
\begin{subfigure}[h]{0.4\textwidth}
\includegraphics[width=0.9\textwidth,trim = {0 0cm 0 0.1cm},clip]{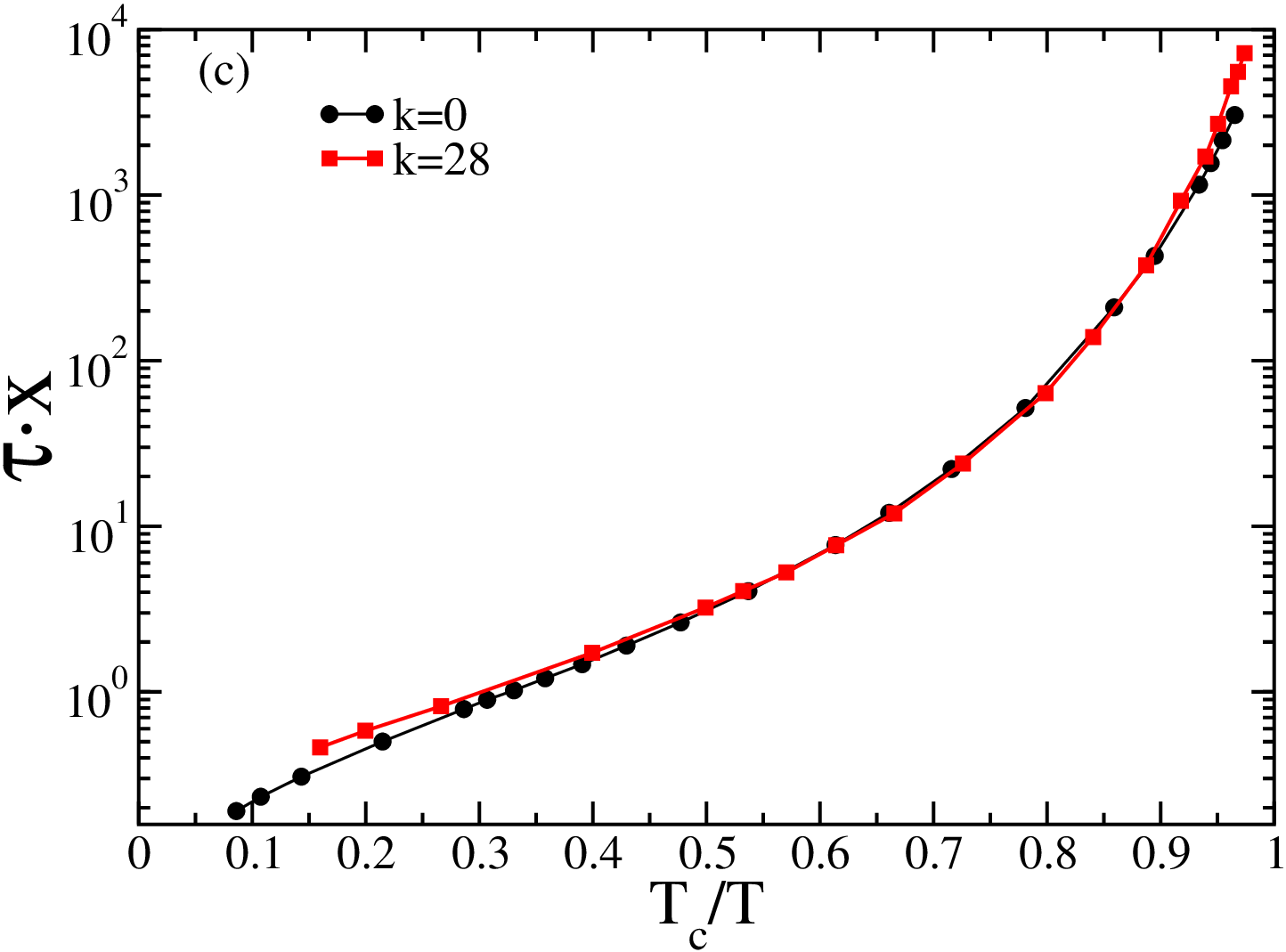}
\end{subfigure}
\caption{(a) The relaxation time obtained from the overlap function as a function of the scaled temperature $(T-T_{c})/T_{c}$ for the $k=0$ and the $k=28$ systems. (b) Same data as in (a) but now with $\tau$ multiplied with a scaling factor $x(k)$. (c) Same data as in (b) as a function of $T_c(k)/T$.
}
\label{fig11_mct_fitting}
\end{figure}

Using this functional form to fit the temperature dependence of the relaxation time we obtain $T_c(k)$ (values are given in Tab.~\ref{table_compare_temp}). In Fig.~\ref{fig11_mct_fitting}(a) we present a log-log plot of the relaxation time as a function of the normalized temperature $(T-T_c)/T_c$. One recognizes that for $k=0$, the increase of $\tau$ with decreasing $T$ is described well by a power law (dashed line), in agreement with previous simulations~\cite{kob-andersen-lett,kob-andersen}. However, at the lowest $T$'s deviations are observed, and the increase in $\tau$ is weaker than the power law predicted by MCT. This deviation is usually attributed to the existence of ``hopping processes'', i.e.~a component in the relaxation dynamics that is not taken into account in the {\it idealized} version of the MCT. The two arrows in the plot delimit the $T$-range in which the power law gives a good description to the data.

For the system with $k=28$ the temperature dependence of $\tau$ is qualitatively very similar to the one for the $k=0$ system, {\it if} one plots the data as a function of the reduced temperature $(T-T_c)/T_c$. The highest temperature at which the data follows the power law (dashed line), marked by an arrow, is around 2$\,T_c$, and very close to the corresponding reduced temperature for the $k=0$ system. However, the lower (reduced) temperature at which $\tau$ starts to deviate from this power law, see arrow, is smaller for the $k=28$ system than the corresponding $T$ for the $k=0$ system, showing that for the former system the mentioned hopping processes are less important, i.e.,~the system is more mean-field like. For the $k=28$ system, this lower limit is about a factor of 3 smaller than the limit for $k=0$; thus the $T$-range in which the idealized MCT can be expected to be reliable has increased significantly by the introduction of the pseudo neighbors. In Tab.~\ref{table_compare_temp} we have also included the value of $T_c$ and one recognizes that the critical temperature for $k=28$ is about 90\% higher than the one for $k=0$, i.e.~the $k$-dependence of $T_c$ is very similar to the one of $T_{\rm onset}$.

According to the analytical calculations for the mean-field $p$-spin model, for which there is no activated dynamics, the onset temperature coincides with the MCT temperature which is also the temperature at which the dynamics diverges \cite{cavagna,kirk_woly1,kirk_woly2}. (Note that this is only true in the thermodynamic limit while for finite systems one has very strong finite-size effects that completely wash out these transitions, see Ref.~\onlinecite{brangian_2001}.) For the GCM it was found that the relative distance between the three temperatures $T_{\rm onset}$, $T_c$, and $T_0$, is much smaller than the one we find here for the $k=0$ system~\cite{kuni-jcp, manoj-gcm}. Thus the reduction of this relative distance with increasing $k$, given in Tab.~\ref{table_compare_temp}, can also be taken as a signature of increasing mean-field like behaviour. 

From Fig.~\ref{fig11_mct_fitting}(a) we recognize that the relaxation times for the $k=28$ system are shorter than the ones for the $k=0$ system if compared at the same reduced temperature. In fact, as plotted in Fig.~\ref{fig11_mct_fitting}(b) on an intermediate time scale the two data sets can be superimposed with high accuracy by applying a multiplicative factor $x(k)$ (see Tab.~\ref{table_compare_temp} for values). Thus we conclude that the main difference in the two data sets is the prefactor $\tau_{\rm MCT}$ in Eq.~(\ref{eq23}). A decrease in $\tau_{\rm MCT}$ implies a faster motion inside the cage, and this is in fact very reasonable since with increasing $k$ the tagged particle is interacting with more particles, thus making its effective cage stiffer. Another way to present this result is to plot the time scale $\tau\cdot x(k)$ as a function of $T_c/T$, see Fig.~\ref{fig11_mct_fitting}(c). We find that this representation of the data gives rise to a collapse of the curves for the different values of $k$, demonstrating that the $T$-dependence is indeed very similar at intermediate temperatures. Hence we conclude that the introduction of the pseudo neighbors does not only increase the $\alpha$-relaxation time strongly but also increase somewhat the attempt frequency with which the particle tries to leave the cage.

\subsection{Wave-vector Dependence of Relaxation Process}

The relaxation time of glass-forming systems depends on the observable considered. Within MCT this dependence is, however, encoded in a prefactor, $\tau_{\rm MCT}$ in Eq.~(\ref{eq23}), while $T_c$ and the exponent $\gamma$ are expected to be independent of the observable. While for many glass-forming systems this is indeed the case, see e.g.~Ref.~\onlinecite{kob-andersen-II}, the present system has at least two relevant length scales, the nearest neighbor distance and the mean distance between the particles and their pseudo neighbors, and hence it is of interest whether the mention factorization works here as well. To probe this we consider the self intermediate scattering function $F_s(q,t)$, where $q$ is the wave-vector~\cite{hansen_mcdonald_86}:

\begin{equation}
F_{s}(q,t) =\frac{1}{N}  \sum_{j=1}^{N} \langle \exp[-i\bf{q}.({\bf r}_j(t)-{\bf r}_i(0))]\rangle \quad .
\end{equation}

\begin{figure}[!bth]
\centering
\begin{subfigure}[h]{0.4\textwidth}
\includegraphics[width=0.9\textwidth,trim = {0 0cm 0 0.1cm},clip]{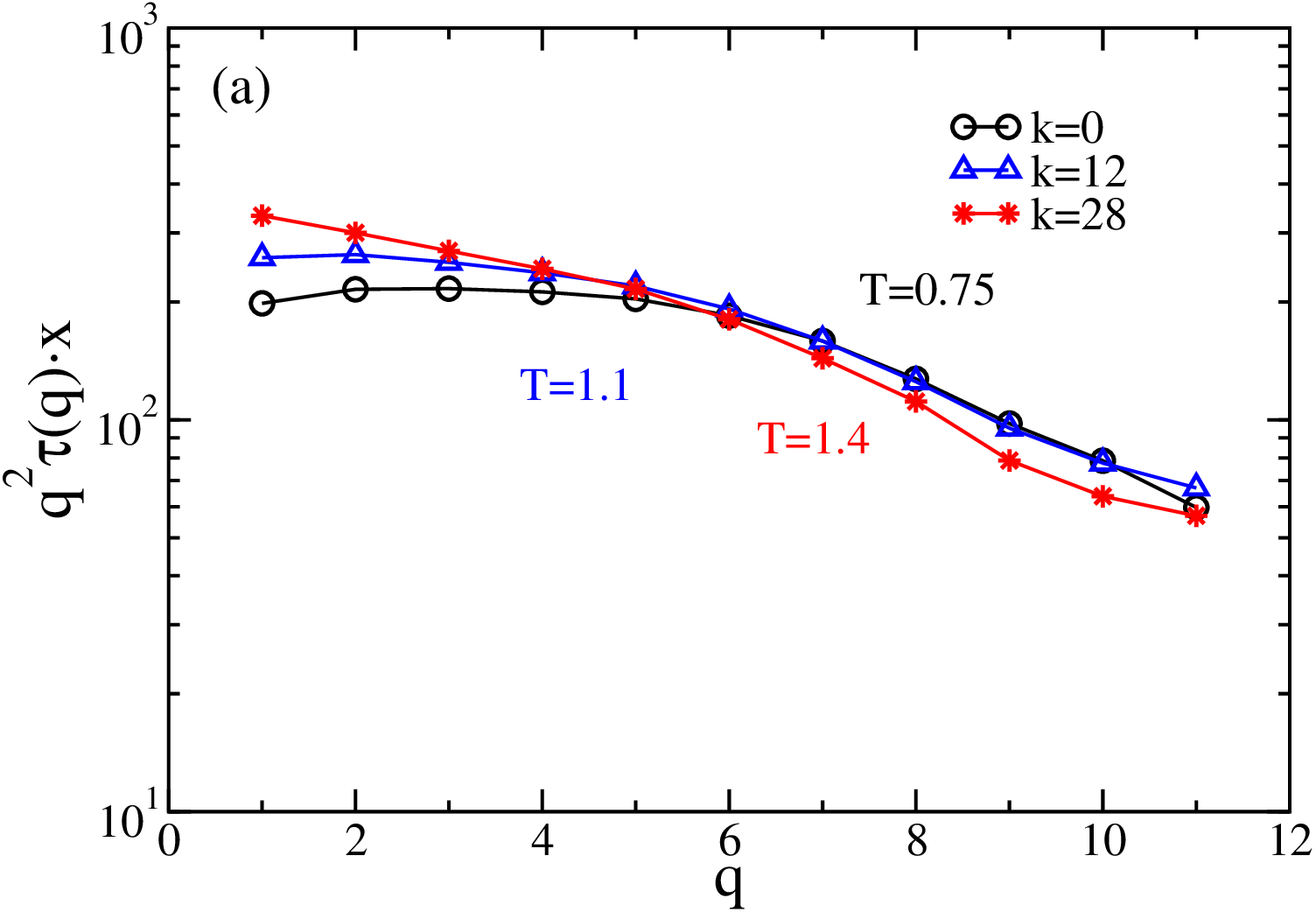}
\end{subfigure}
\begin{subfigure}[h]{0.4\textwidth}
\includegraphics[width=0.9\textwidth,trim = {0 0cm 0 0.1cm},clip]{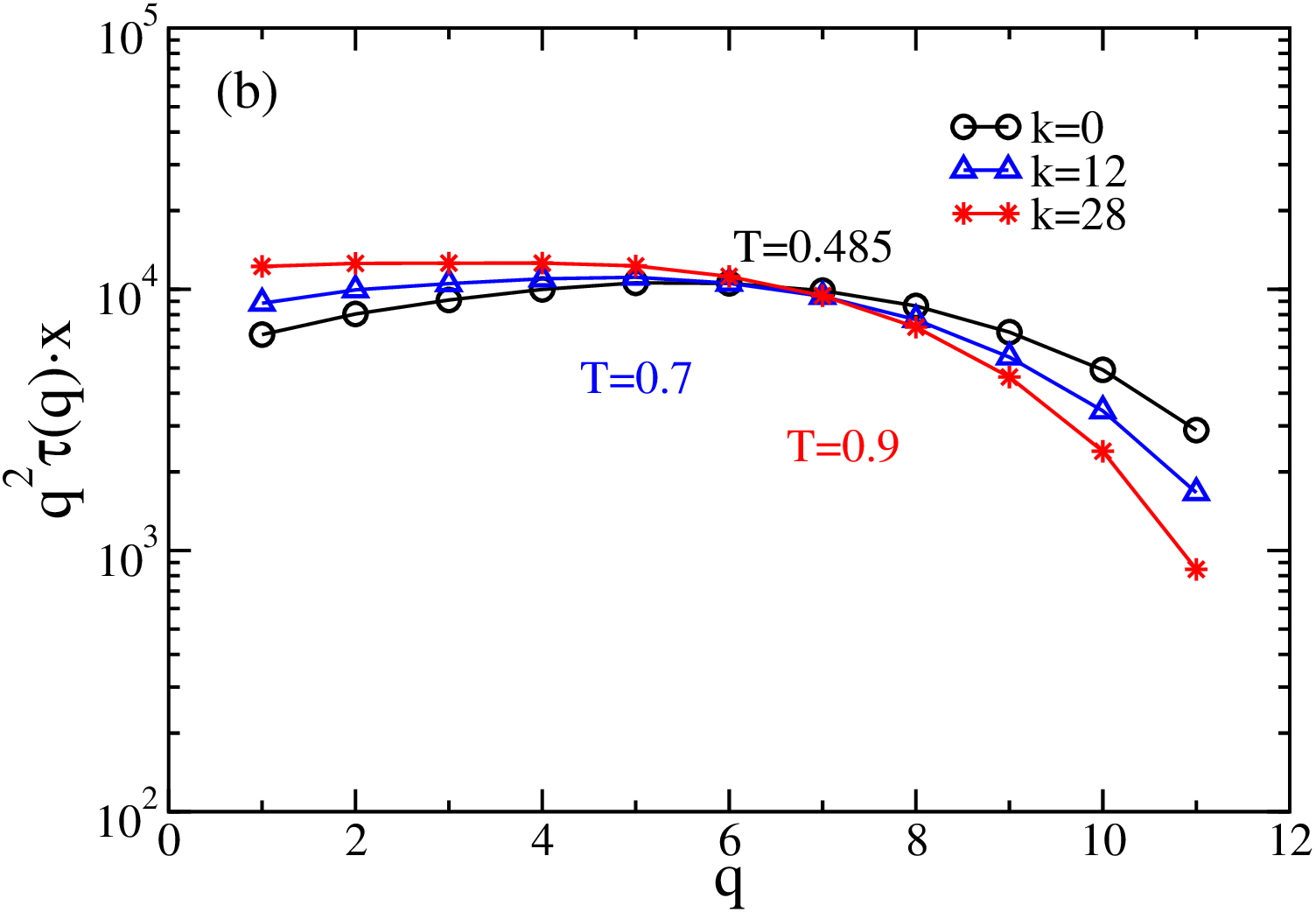}
\end{subfigure}
\caption{$q^{2}\tau(q) \cdot x(k)$ as a function of the wave-vector $q$. Panels (a) and (b) are for two different reduced temperatures.  The values of $x(k)$ are given in Table~\ref{table_compare_temp}}.

\label{fig_q_square_tau}
\end{figure}

We define the relaxation time $\tau(q)$ via $F_s(q,\tau(q)))=1/e$ and thus can study its dependence on the length scale. In Fig.~\ref{fig_q_square_tau} we show the $q$-dependence of $\tau(q)$ for three values of $k$. Since one expects that at small wave-vectors $\tau(q)$ is proportional to $q^{-2}$, i.e.~the hydrodynamic behavior, we plot directly $q^2\tau(q)$. Panel (a) is for a fixed reduced temperature slightly below the onset temperature while panel (b) corresponds to a significantly supercooled state. In the context of Fig.~\ref{fig11_mct_fitting}(b) we have seen that, at a fixed reduced temperature, the relaxation time $\tau$, obtained from the decay of the overlap function, shows a weak dependence on $k$, leading to the introduction of the factor $x(k)$. In order to take into account this $k$-dependence we have multiplied also in Fig.~\ref{fig_q_square_tau} the relaxation times $\tau(q)$ with the {\it same} factor $x(k)$. The graphs shows that for $q\approx 6.5$, i.e.~close to the peak of the static structure factor, the relaxation times for the different systems coincide perfectly, which demonstrates that for this 
wave-vector the overlap and $F_s(q,t)$ probe the same type of dynamics. For the other wave-vectors 
considered, the $\tau(q)$ curves for the different systems show a $q-$dependence that depends on $k$, but this dependence is relatively weak. Hence we conclude that the presence of the pseudo neighbors does not introduce a new length scale that influences the relaxation dynamics in a significant manner.


\subsection{Dynamic Heterogeneity}
\begin{figure}[!bth]
\centering
\begin{subfigure}[h]{0.4\textwidth}
\includegraphics[width=0.9\textwidth,trim = {0 0cm 0 0.1cm},clip]{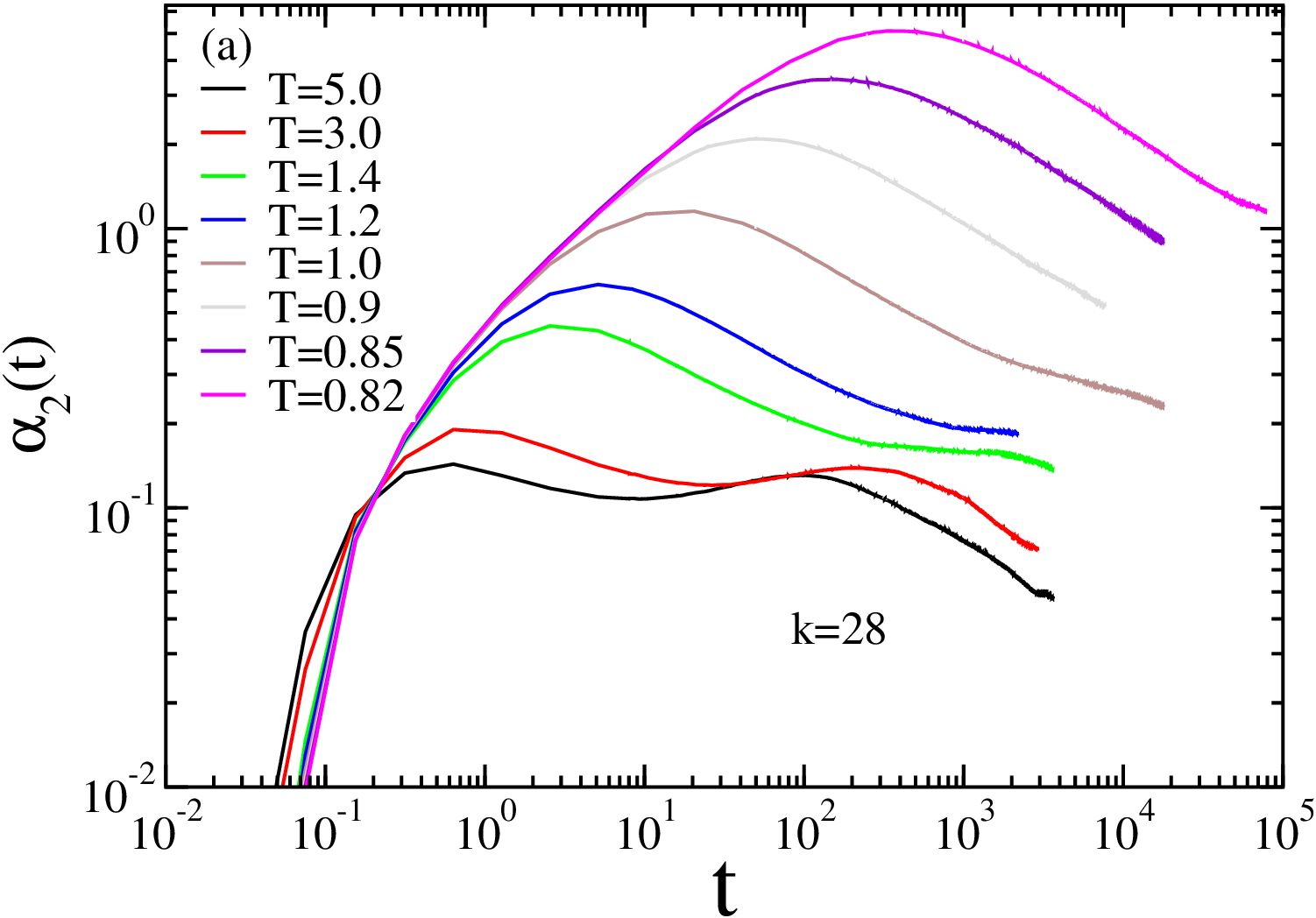}
\end{subfigure}
\begin{subfigure}[h]{0.4\textwidth}
\includegraphics[width=0.9\textwidth,trim = {0 0cm 0 0.1cm},clip]{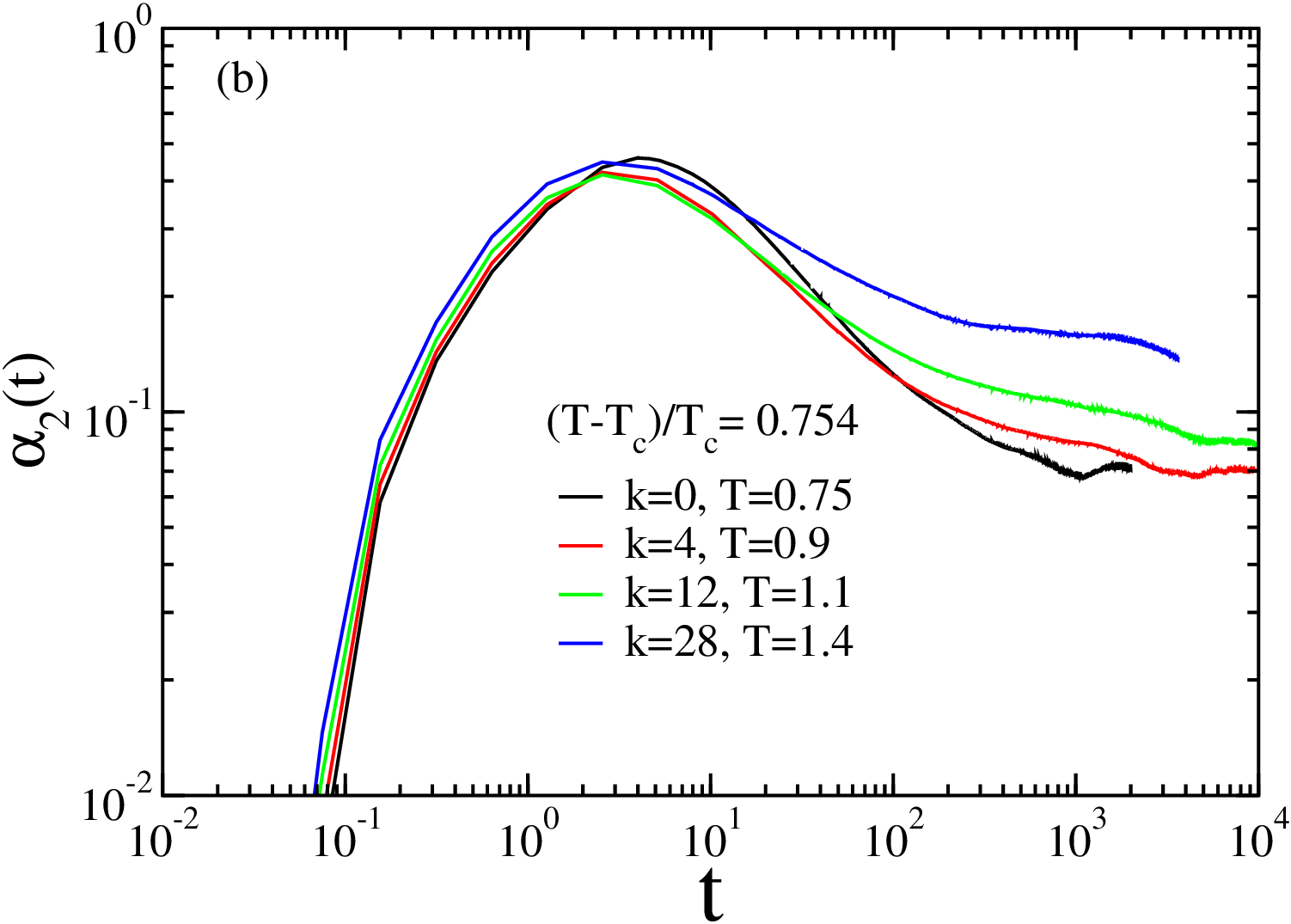}
\end{subfigure}
\caption{(a) The time dependence of the non-Gaussian parameter, $\alpha_{2}$, at different temperatures for the $k=28$ system. $\alpha_{2}(t)$ shows a double peak structure. (b) $\alpha_2(t)$ at fixed reduced temperature and different values of $k$. The peak at short times is independent of $k$ while the one at long times grows with increasing $k$.
}
\label{fig12_alpha2_vs_T}
\end{figure}

One of the hallmarks of glassy dynamics is that time correlation functions are stretched in time. The reason for this non-Debye relaxation has been a long-standing puzzle with the contrasting views that each small domain of the sample shows the same stretched time dependence or, alternatively, that the stretching is related to dynamical heterogeneities~\cite{ediger_review}. Experiments and simulations have shown that the homogeneous scenario is not compatible with the observations, i.e.~glass-forming systems do have a significant amount of dynamical heterogeneities (DH)~\cite{donati_prl,kob_prl,heuer,hurley_harrowell,kim-saito}. In this final section, we therefore discuss the $k$-dependence of these DH and probe whether with increasing $k$ one does indeed find a decrease of these fluctuations, the behavior expected for a mean-field system.

\begin{figure}[!th]
\includegraphics[width=0.45\textwidth,trim = {0 0cm 0 0.1cm},clip]{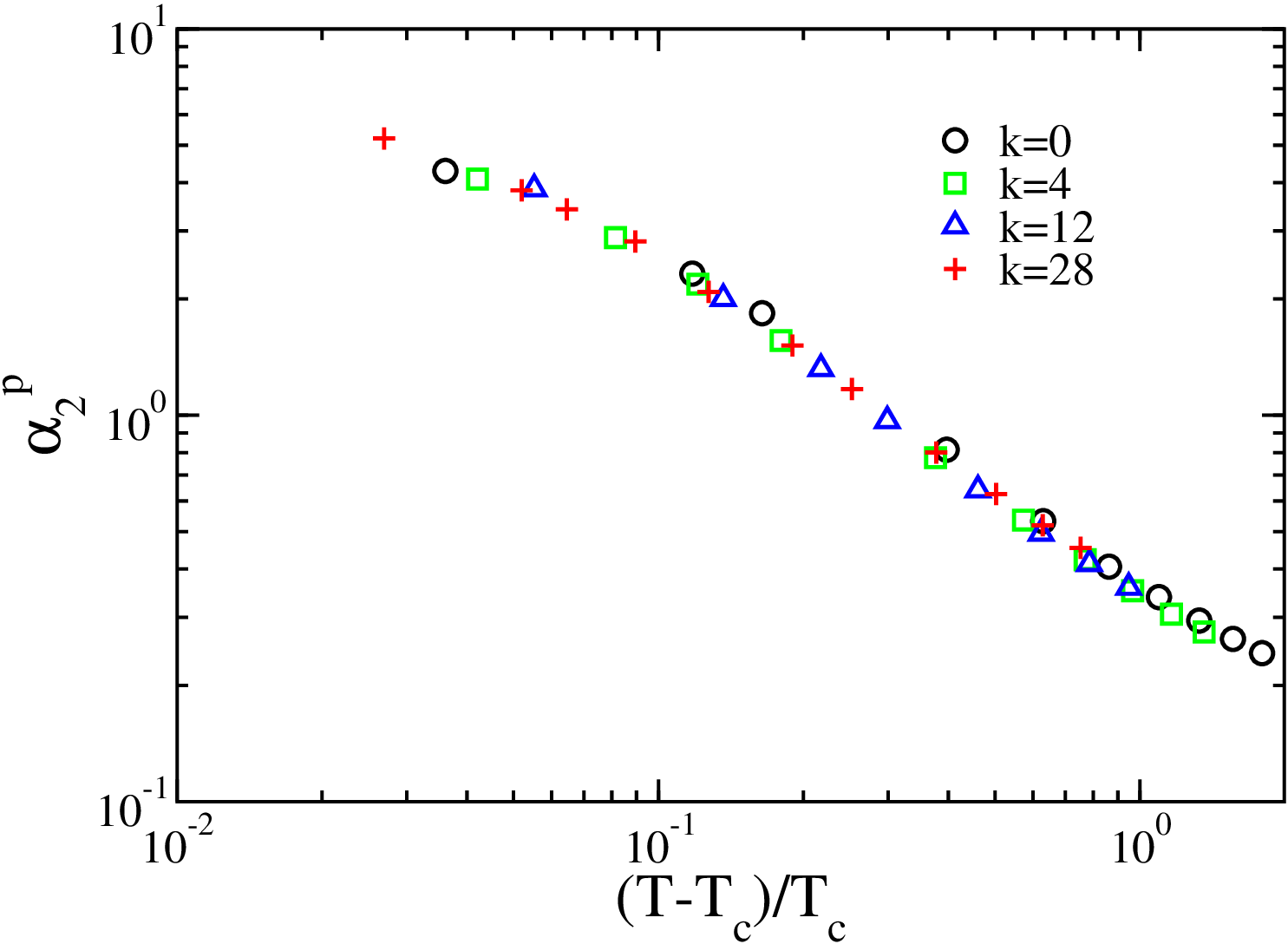}
\caption{The peak height of $\alpha_{2}$ as a function of the reduced temperature $(T-T_c)/T_c$ for different values of $k$.
}
\label{fig13_alpha2_same_epsilon}
\end{figure}

One first step to probe the DH is to look at the so-called non-Gaussian parameter (NGP) $\alpha_2(t)$ which is define by

\begin{equation}
\alpha_{2}(t)=\frac{3\;\langle r^4(t)\rangle}{5\;\langle r^2 (t)\rangle^{2}}-1 \quad ,
\label{eq25a}
\end{equation}

\noindent
where $r(t)$ is the displacement of a tagged particle within a time $t$. Thus $\alpha_2(t)$ measures whether or not the distribution of the particle displacement is Gaussian~\cite{odagaki,kob-andersen,donati_prl,sastry_jack_star}.

In Fig.\ref{fig12_alpha2_vs_T}(a) we plot the NGP for the $k=28$ system. Interestingly one finds that at high temperatures $\alpha_2(t)$
has {\it two} peaks: A first one at $t$ around 0.6 and a second one at $t\approx 150$. The first time is close to the timescale at which the MSD crosses over from the ballistic regime to the diffusive one and thus corresponds to the start of the NN-$\alpha$-process, in agreement with earlier studies~\cite{kob-andersen}. The second peak has so far not been seen in the glass-forming systems considered before and is likely due to the breaking of the bonds with the pseudo neighbors, i.e.~the PN-$\alpha$-relaxation. Note that the presence of this second peak is coherent with our findings for the MSD, see Fig.~\ref{fig9_msd_slope}(c), for which we observed a plateau in the slope that, for $T=2.0$, ended at around $t=10^2$ and we had argued that this is due to the motion of the pseudo neighbors. If $T$ is lowered, the first peak in $\alpha_2(t)$ rises quickly and dominates the second peak, i.e.~on overall the time dependence of the NGP becomes again quite similar to the one that has been observed in previous studies of glass-forming systems. 
The main difference is that in our case the second peak will make the decay of $\alpha_2(t)$ slow since at long times the dynamics will be influenced by the pseudo neighbors, which decorrelate only slowly (see the data for the MSD in Fig.~\ref{fig9_msd_slope}).

The influence of the pseudo neighbors on $\alpha_2(t)$ is shown in Fig.~\ref{fig12_alpha2_vs_T}(b) where we plot this function for different values of $k$ but keeping $(T-T_c)/T_c$ constant. One sees that at short and intermediate times, i.e.~around the peak, the curves are independent of $k$, which shows that the NN-$\alpha$-process is not affected by the presence of the pseudo-neighbors. Only at longer times, the curves for large $k$ are higher than the ones for small $k$, showing that the pseudo neighbors affect the NGP only at time scales that are beyond the time scale of the first maximum in the NGP. Since with decreasing temperature the peak corresponding to the NN-$\alpha$-relaxation grows quicker than the second peak we can conclude that the dominant feature in $\alpha_2(t)$ is due to the NN-$\alpha$-process, except if $k$ becomes much larger than the values we consider here.

\begin{figure}[!ht]
\includegraphics[width=0.45\textwidth]{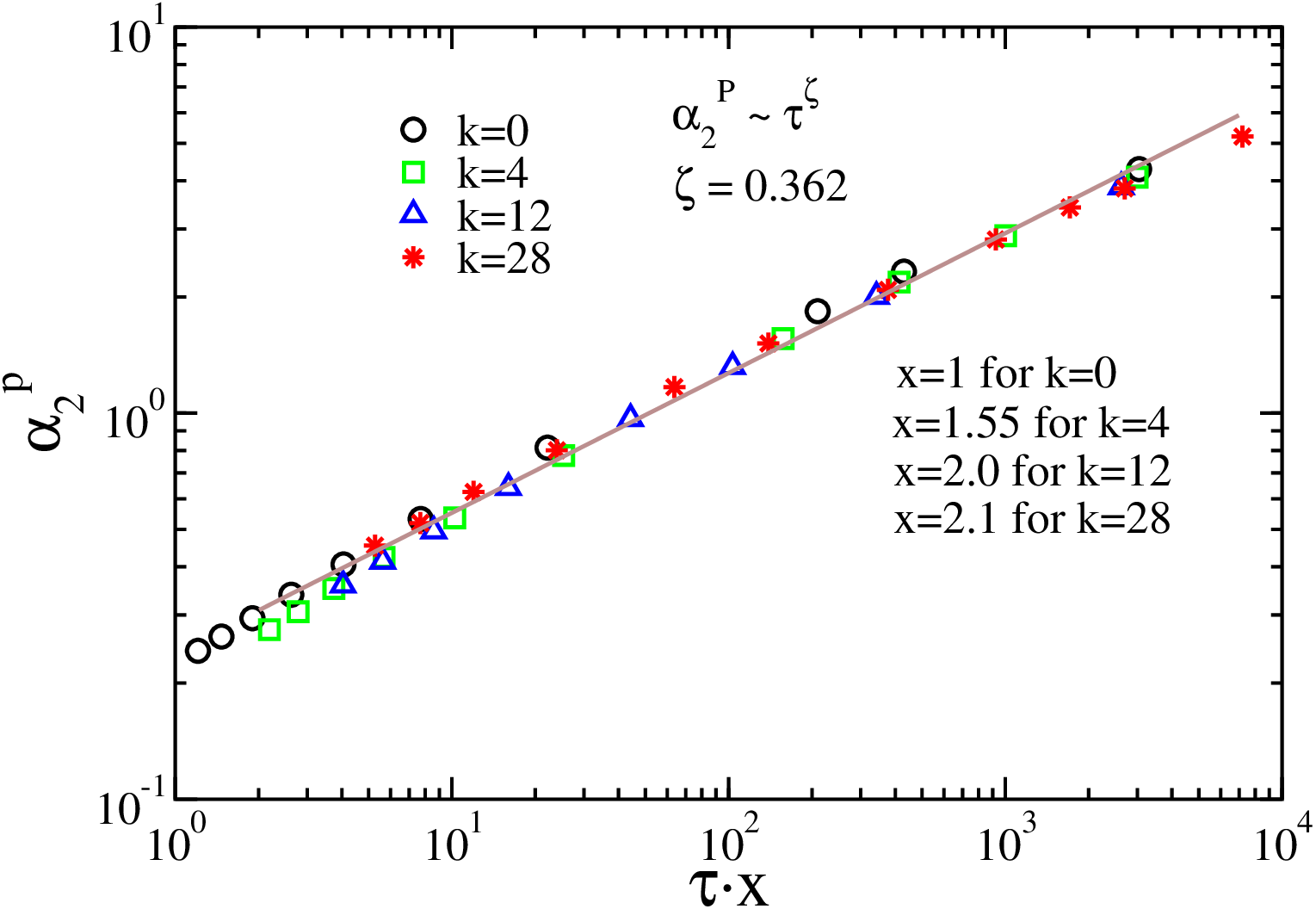}
\caption{The peak height of $\alpha_{2}$ as a function of the $\alpha$-relaxation time $\tau$ multiplied by $x(k)$ for different values of $k$. Also included is a fit to the data with a power law.
}
\label{fig14_alpha2_with_tau}
\end{figure}

\begin{figure}[!bt]
\includegraphics[width=0.45\textwidth]{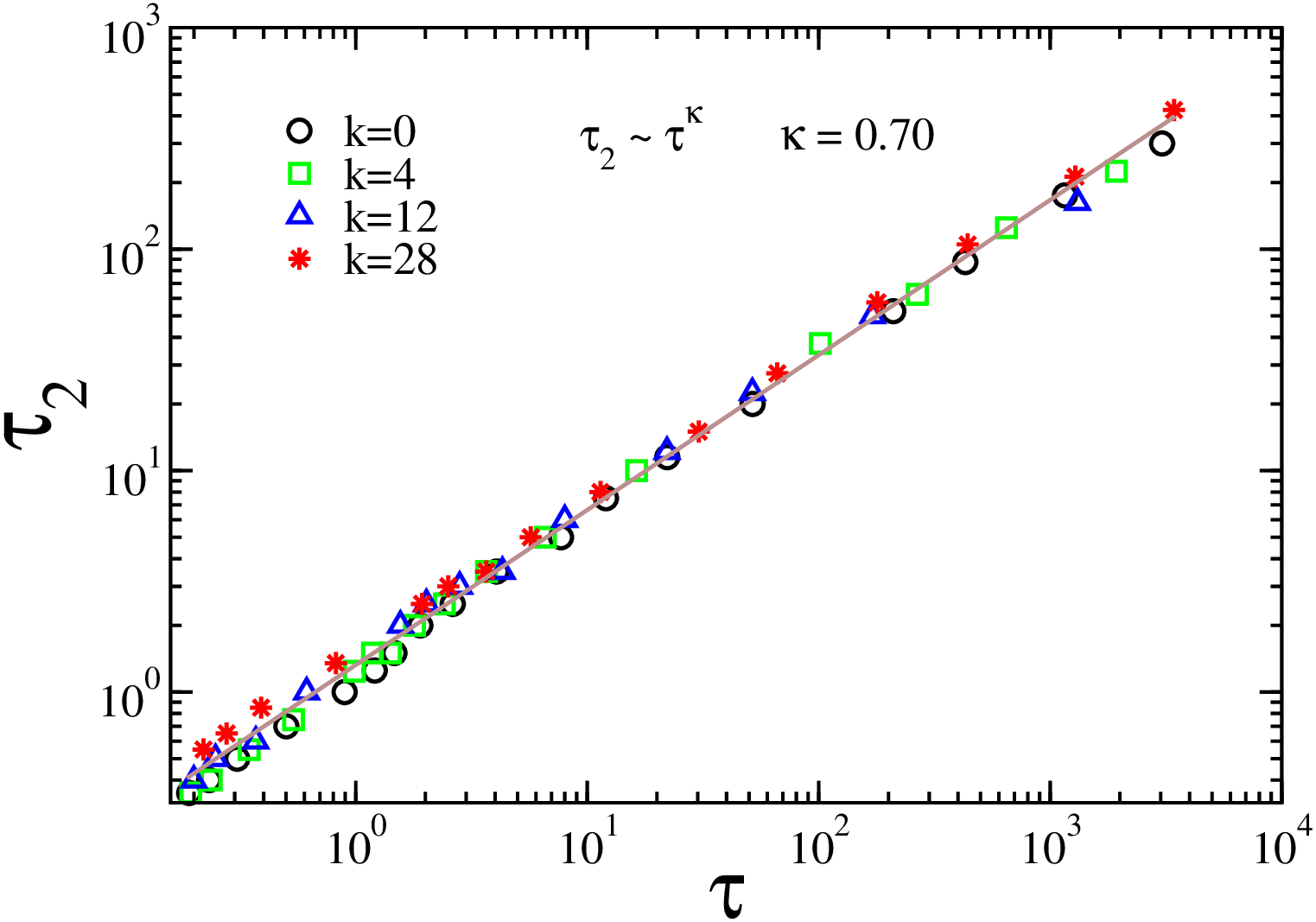}
\caption{
The time scale
$\tau _{2}$ at which $\alpha_2(t)$ peaks, as a function of the $\alpha$-relaxation time $\tau$. The solid line is a power law with an exponent $\kappa=0.70$.
}
\label{fig15_tau2_vs_tau}
\end{figure}

In Fig.~\ref{fig13_alpha2_same_epsilon} we show $\alpha_2^p$, the height of the peak in $\alpha_2(t)$, as a function of the reduced temperature $(T-T_c)/T_c$.
Surprisingly we find that this quantity is completely independent of $k$, i.e.~the strength of the non-Gaussianity of the relaxation dynamics does not depend on whether or not the system is mean-field like. In other words, the statistics of the displacement of a tagged particle is independent of the number of pseudo neighbors, if measured at the same reduced temperature. This result reflects the fact that the first peak in $\alpha_2(t)$ is dominated by the dynamics in which the tagged particle leaves the cage formed by its nearest neighbors.

\begin{figure}[!bt]
\includegraphics[width=0.45\textwidth]{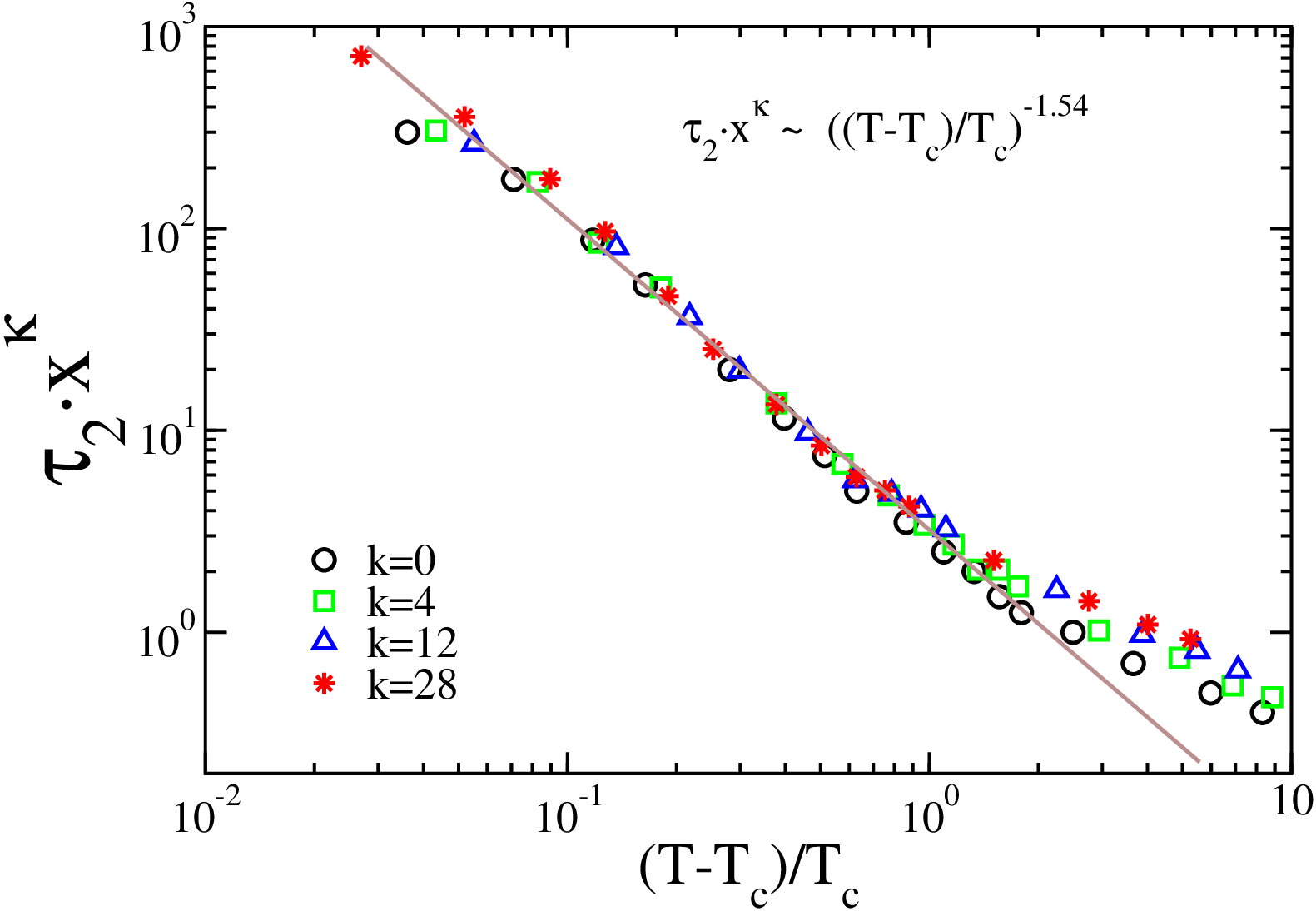}
\caption{$\tau _{2} x(k)^\kappa$
as a function of the reduced temperature  $(T-T_{c})/T_{c}$.  The solid line is a power law with exponent -1.54.
}
\label{fig15_tau2_epsilon}
\end{figure}

Note that $\alpha_2^p$ shows a bend at around $(T-T_c)/T_c\approx 0.1$. Although we did not investigate the origin of this change in the $T$-dependence, we expect it to be the signature of the onset of the hopping processes mentioned above. The bend indicates that these processes start to become prominent at around 10\% above $T_c$, a value that seems to be coherent with the observation from Fig.~\ref{fig11_mct_fitting} regarding the $T$-dependence of the relaxation times.

One might wonder whether the master curve in Fig.~\ref{fig13_alpha2_same_epsilon} is just due to the choice of the scaling factor of the temperatures, i.e.~$T_c$. To test this possibility, we show in
Fig.~\ref{fig14_alpha2_with_tau} the same data as a function of the relaxation time $\tau$ multiplied by the same factor $x(k)$ that was used to obtain a master curve in Fig.~\ref{fig11_mct_fitting}(b). We recognize that this representation leads to a very nice collapse of the data onto a master curve which, for intermediate and long relaxation times, can be described well with a power law with an exponent close to 0.36 (see solid line in the figure). It is remarkable that the hopping processes discussed above, which lead to the bends in the different curves if the temperature approaches $T_c$, do not seem to affect the validity of the power law. 
At present, it is not clear up to which value of $\tau$ this power law will hold, in particular, whether it will be observed at temperatures below $T_c$. Future studies on this point will certainly be of interest to understand better the relaxation dynamics of glass-forming liquids.  

In Fig.~\ref{fig15_tau2_vs_tau} we plot $\tau_2$, the time at which $\alpha_2(t)$ peaks, as a function of the $\alpha$-relaxation time $\tau$. Surprisingly we find that the two quantities show a simple relation with each other in the form of a power law with an exponent $\kappa=0.70$ (solid line). This result can be rationalized within the framework of MCT as follows: $\alpha_2(t)$ is related to the shape of the self part of the van Hove function in that it measures its deviation from a Gaussian~\cite{odagaki,kob-andersen}. At the end of the caging regime, i.e.~the $\beta$-relaxation, some of the particles will have already left their cage, thus giving rise to a tail to the right of the main peak of the van Hove function. It is this tail that is responsible for the non-Gaussian shape of the van Hove function and hence leads to an increase of $\alpha_2(t)$. Thus it is reasonable to assume that $\tau_2$ is directly related to the time scale of the $\beta$-relaxation $\tau_\beta$. MCT predicts that the latter time scale increases like 

\begin{equation}
\tau_\beta \propto (T-T_c)^{-1/(2a)} \quad .
\label{eq29}
\end{equation}

The $\alpha$-relaxation time $\tau$ is instead predicted by MCT to increase like

\begin{equation}
\tau \propto (T-T_c)^{-1/(2a)-1/(2b)} = (T-T_c)^{-\gamma} \quad .
\label{eq30}
\end{equation}

In Eqs.~(\ref{eq29}) and (\ref{eq30}) the parameters $a$ and $b$ can in principle be calculated from the $T$-dependence of the static structure factor or, exploiting Eq.~(\ref{eq30}), determined from the $T$-dependence of the relaxation time~\cite{gotze,binder_kob_book,buchalla_gotzess}. For the $k=0$ system it has been found that $a$ is around $0.324$ and $b$ is around $0.627$ ~\cite{buchalla_gotzess,kob-andersen,gleim_prl,nauroth}. Combining these last two equations gives, under the assumption that $\tau_2 \propto \tau_\beta$, 

\begin{equation}
\tau_2 \propto \tau^{b/(a+b)}\quad .
\label{eq31}
\end{equation}

\noindent
Thus we find a power law dependence with an exponent of $0.66$ (using the mentioned values of $a$ and $b$), which is indeed very close to our exponent $\kappa$ from the fit (0.7). We mention here that the observed power law extends over the whole accessible range of $\tau$, i.e.~it also includes the temperature regime in which we expect hopping processes to be present. 
To the best of our knowledge this simple connection between $\tau_2$ and $\tau$ has not been reported before. Since, however, we find it to hold for all values of $k$, we expect it to be valid for other glass-forming systems as well and hence it will be of interest to check this in the future.

To get Eq.~(\ref{eq31}) we have made the assumption that $\tau_2$ is proportional to $\tau_\beta$. As argued above, this hypothesis is reasonable since it can be expected that the non-Gaussian parameter peaks at a time at which a substantial number of particles start to leave their cage and MCT defines $\tau_\beta$ as the time at which the correlator starts to drop below the plateau at intermediate times~\cite{gotze_book}. 
Previous studies have therefore made the assumption that $\tau_\beta$ can be determined from the minimum in the slope of the MSD~\cite{rajsekhar}. However, we argue that such an identification might be misleading: For the case of a system with Newtonian dynamics, the phonons that govern the short-time dynamics mask the critical decay of the time correlation functions thus also masking the correlation between the above-mentioned minimum and $\tau_\beta$. (This effect is, however, absent if the system has a Brownian dynamics~\cite{gleim_prl}.) Therefore we think it is more appropriate to determine $\tau_\beta$ from a quantity that is not directly influenced by these vibrational modes, such as the $\alpha_2(t)$ considered here. In Fig.~\ref{fig9_msd_slope}(a)-(c) we have also included for the various curves the times $\tau_2$, arrows pointing upward, and one sees that they do not correspond to the location of the minimum in the curves but that they are located at somewhat larger times, as expected because of the mentioned effect of the phonons. Although at present we do not have any solid proof why $\tau_2$ does indeed correspond to $\tau_\beta$, our finding that the relation between $\tau_2$ and $\tau$ given by Eq.~(\ref{eq31}) is obeyed by our data does speak in favour of this identification. More tests on this using a system with Brownian dynamics would certainly be useful to clarify this point further.

\begin{figure}[!th]
\centering
\begin{subfigure}[h]{0.4\textwidth}
\includegraphics[width=0.9\textwidth]{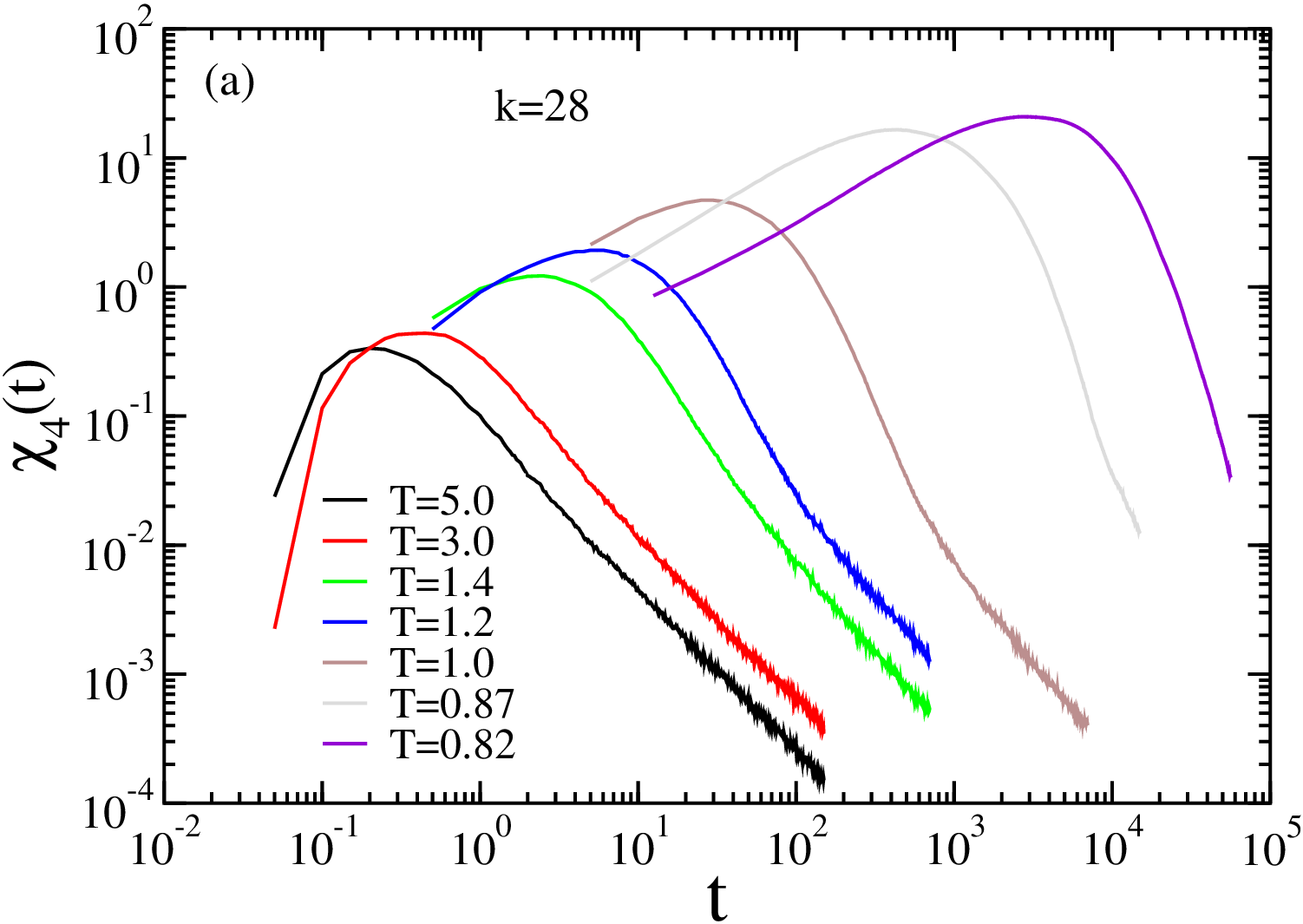}
\end{subfigure}
\begin{subfigure}[h]{0.4\textwidth}
\includegraphics[width=0.9\textwidth,trim = {0 0cm 0 0.1cm},clip]{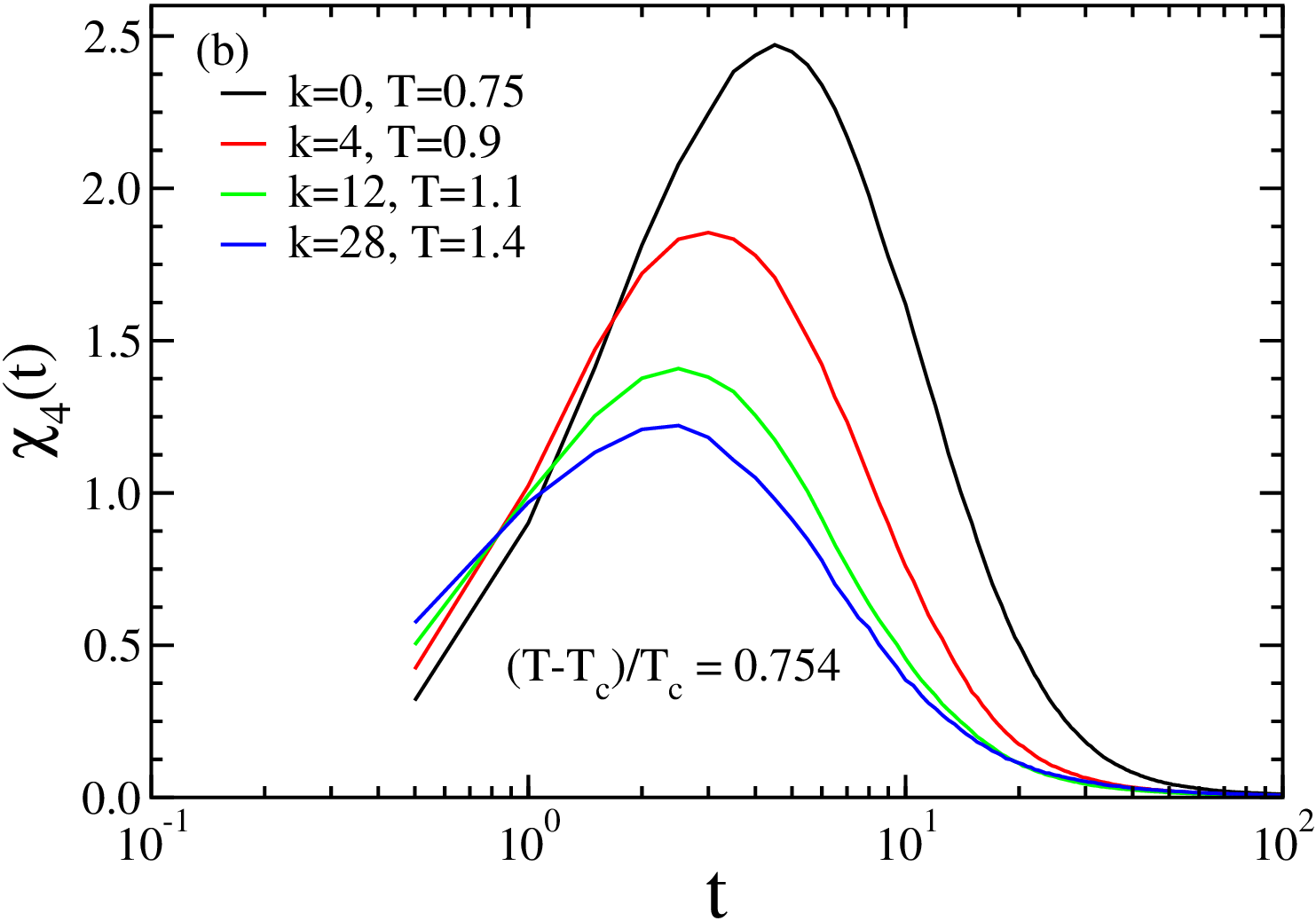}
\end{subfigure}
\caption{(a) The time dependence of the dynamical susceptibility $\chi_{4}(t)$ for different temperatures for the  $k=28$ system. $\chi_{4}(t)$ increases with decreasing temperature. 
(b) Time dependence of $\chi_4$ at a fixed reduced temperature $(T-T_c)/T_c$ for different values of $k$.
}
\label{fig16_chi4_vs_T}
\end{figure}

Finally we show in Fig.~\ref{fig15_tau2_epsilon} the time at which $\alpha_2(t)$ peaks, $\tau_2$, as a function of $(T-T_c)/T_c$. Since we have argued in the context of Fig.~\ref{fig11_mct_fitting} that the $k$-dependence of $\tau$ will include a factor $x(k)$ that is related to the short time dynamics, and we also showed that $\tau_{2} \propto \tau^{\kappa}$ (Fig.~\ref{fig15_tau2_vs_tau}), we plot directly $\tau_2 \cdot x(k)^{\kappa}$, with the values of $x(k)$ obtained from Fig.~\ref{fig11_mct_fitting} and $\kappa$ from Fig.~\ref{fig15_tau2_vs_tau}.
We recognize that the data for the different values of $k$ fall nicely on a master curve which follows a power law with an exponent around -1.54. Also this result can be understood within the framework of MCT since Eq.~(\ref{eq29}) predicts that the slope should be given by $-1/(2a)$ which for $a=0.324$ results in an exponent of $-1.54$, in excellent agreement with the data from the fit in Fig.~\ref{fig15_tau2_epsilon}.

Next we discuss the other parameter which is often related to the dynamic heterogeneity, the dynamic susceptibility. The fluctuations of the overlap function $Q(t)$ are related to a dynamic susceptibility which indicates whether or not the system relaxes in a cooperative manner, i.e. shows dynamical heterogeneities~\cite{berthier_jcp_2007_1,berthier_jcp_2007_2,coslovich_ozawa_kob}. Thus one defines

\begin{equation}
\chi_{4}(t)=\frac{1}{N}\big[\langle Q^{2}(t)\rangle-\langle Q(t)\rangle^{2}\big]
\label{eq25b}
\end{equation}

\noindent
as a measure to quantify this cooperativity. In Fig.~\ref{fig16_chi4_vs_T}(a) we show the time dependence of $\chi_4$ for the system with $k=28$ at different temperatures. In agreement with earlier studies,\cite{sastry-SE}, we find that $\chi_4$ shows a marked peak the height of which increases with decreasing temperature and also its position shifts to larger times upon decreasing $T$, i.e.~the cooperativity becomes more pronounced and occurs at later times.  In panel~(b) of the figure we present $\chi_4$ for different values of $k$ while keeping the normalized temperature $(T-T_c)/T_c$ constant. The graph demonstrates that with increasing $k$ the height of the peak decreases quickly, indicating that the system does indeed become more mean-field like, as expected, and in agreement with previous simulations of mean-field like models~\cite{mari-kurchan,sastry-SE}. This $k$-dependence is thus very different from the one seen for the height of the peak in $\alpha_2$, highlighting the difference between the two quantities, despite their (apparently) similar time dependence. We also note that with increasing $k$ the location of the peak in $\chi_4(t)$ shifts to shorter times, in qualitative agreement with the fact that, at fixed reduced temperature, the $\alpha$-relaxation time decreases somewhat, see Fig.~\ref{fig11_mct_fitting}(a).

\begin{figure}[!bth]
\centering
\begin{subfigure}[h]{0.45\textwidth}
\includegraphics[width=0.9\textwidth,trim = {0 0cm 0 0.1cm},clip]{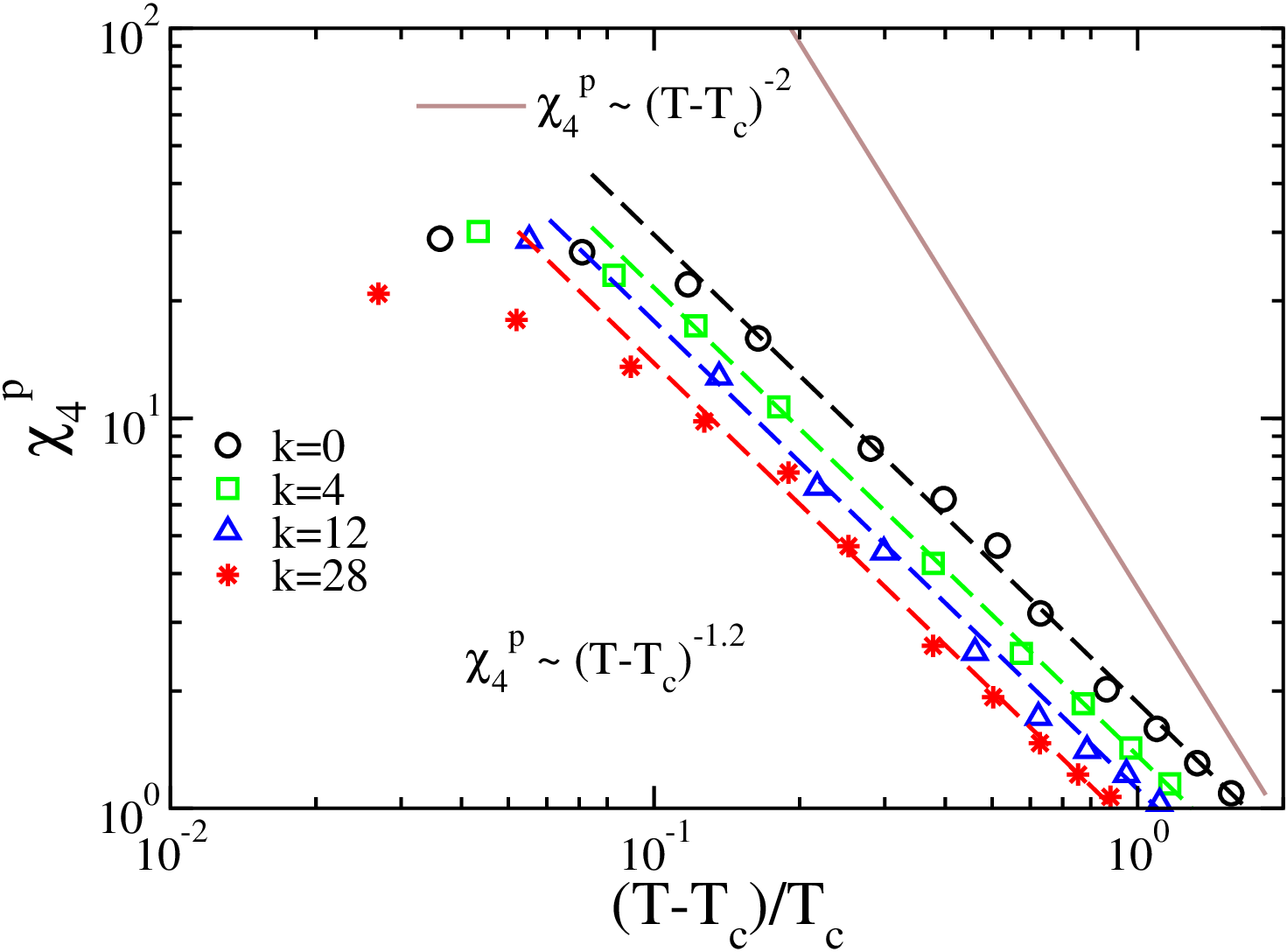}
\end{subfigure}
\caption{Height of the peak in $\chi_{4}(t)$ as a function of the  reduced temperature for different values of $k$. The dashed lines are power laws with exponent -1.2 and the solid line is a power law with an exponent -2.
}
\label{fig17_chi4_same_epsilon}
\end{figure}

To probe in more detail how the height of the peak in $\chi_4(t)$, $\chi_4^p$, depends on $T$ and $k$ we show in Fig.~\ref{fig17_chi4_same_epsilon} this height as a function of the reduced temperature. We see immediately that this representation of the data does not give rise to a master curve. With increasing $k$, the curves move downwards, a $k$-dependence that is in contrast to the one we found for $\alpha_2^p$ shown in Fig.~\ref{fig13_alpha2_same_epsilon}. Thus we conclude that with increasing $k$ the dynamical heterogeneities decrease, i.e.~the system becomes more mean-field like. However, we point out that even in the mean-field limit these heterogeneities cannot be expected to vanish completely~\cite{mari-kurchan,brangian_jpa_2002} which shows that this aspect of the dynamics is a delicate feature that is highly non-trivial.

From the figure, one can conclude that for reduced temperatures higher than around 0.1 the height of the peak shows a power law dependence on the reduced temperature and we find an exponent of -1.2 that is independent of $k$, which implies that the dependence of $\chi_4^p$ on the number of pseudo neighbors is encoded in the prefactor of the power law. 

The presence of power laws in $\chi_4^p$ can be rationalized by means of MCT. This theory predicts that the dynamical susceptibility in the $NVT$ ensemble is given by

\begin{equation}
\chi_4^{\rm NVT}(t) = \chi_4^{\rm NVE}(t)+\frac{T^2}{c_V}\left(\frac{dQ(t)}{dT}\right)^2 \quad,
\label{eq25c}
\end{equation}

\noindent
where $c_V$ is the specific heat at constant volume~\cite{berthier_jcp_2007_1,berthier_jcp_2007_2,coslovich_ozawa_kob}. Evaluating this expression at $t=\tau$, thus giving the height of the peak, $\chi_4^p$,  one finds that the first term on the right-hand side of the equation increases like $(T-T_c)^{-1}$ while the second one is found to be proportional to $(T-T_c)^{-2}$. Hence the power law with exponent -1.2 we find at intermediate and higher temperatures can be interpreted to be 
due to the power law from the first term, i.e.~with an exponent -1.0, which is somewhat augmented by the presence of the second term, thus giving rise to a power law with an effective exponent smaller than -1. Thus if the mentioned hopping processes would be absent one would expect that at sufficiently low temperatures, the power law crosses over to one with an exponent -2. Whether this is indeed the case will have to be tested for systems in which one is able to suppress these hopping processes, a work that is left for the future.

\begin{figure}[!bth]
\centering

\begin{subfigure}[h]{0.45\textwidth}
\includegraphics[width=0.9\textwidth,trim = {0 0cm 0 0.1cm},clip]{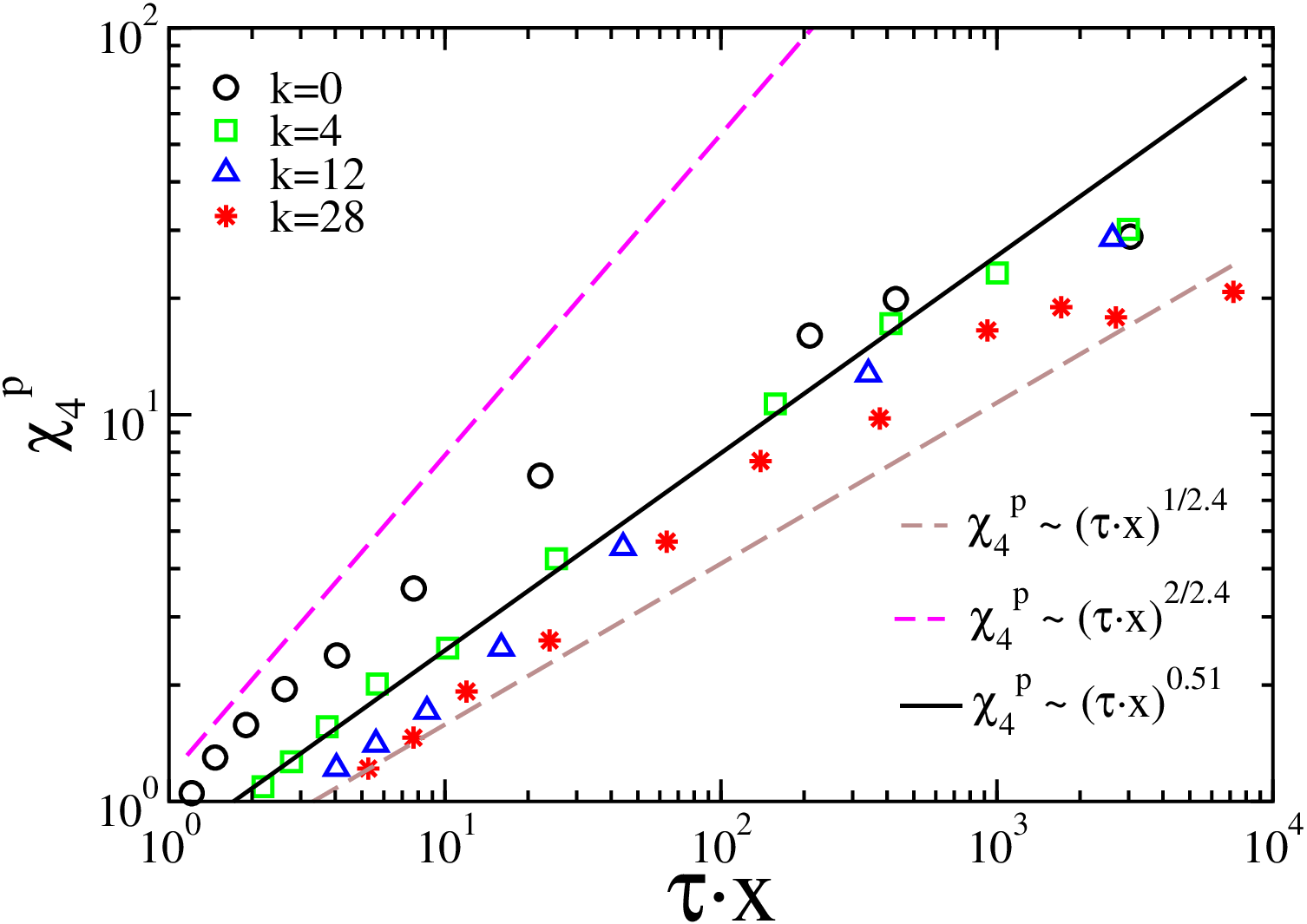}
\end{subfigure}
\caption{The height of the peak in $\chi_{4}$ as a function of $\tau \cdot x(k)$ for different values of $k$. The solid line is a power law fit to the data for $k=4$. The two dashed lines are power laws with exponents that correspond to the theoretical upper and 
lower bounds from Eq.~(\ref{eq26}).
}
\label{fig18_chi4_with_tau}
\end{figure}

Since the representation of the data in Fig.~\ref{fig17_chi4_same_epsilon} depends on the choice of $T_c$, it is also useful to look at the $k$-dependence of $\chi_4^p$ in a more direct manner. This is done in Fig.~\ref{fig18_chi4_with_tau} where we plot this quantity as a function of the $\alpha$-relaxation time $\tau$. (Also here we use $\tau \cdot x(t)$ as abscissa, in order to take into account the trivial $k$ dependence of the relaxation time.) We see that the shape of the curves for the different $k$ is basically independent of $k$, but that the absolute value of $\chi_4^p$ at fixed $\tau \cdot x(k)$ decreases with increasing $k$. (The same conclusion is reached if one uses just $\tau$ as the abscissa.) Hence we confirm the conclusion from Fig.~\ref{fig16_chi4_vs_T}(b) that the heterogeneity of the system decreases with increasing $k$. For small and intermediate values of $\tau$, the data falls approximately on a straight line, and a power law fit gives an exponent 0.51 (solid line). Expressing the $T$-dependence on the right hand side of Eq.~(\ref{eq25c}) as a function of $\tau=(T-T_c)^{-\gamma}$, see Eq.~(\ref{eq23}), we
obtain for the height of the peak

\begin{equation}
\chi_4^p = A \tau^{1/\gamma}+B \tau^{2/\gamma} \quad ,
\label{eq26}
\end{equation}

\noindent
where $A$ and $B$ are expressions that have only a weak $T$-dependence. Using our value  $\gamma=2.4$ gives for the exponent of the first and second term 0.42 and 0.83, respectively. These values are thus upper and lower bounds (included in Fig.~\ref{fig18_chi4_with_tau} as well) and the exponent we extract from our data, 0.51, is thus not too far from the lower limit. So, although our data do not allow to make strong statements about the validity of Eq.~(\ref{eq26}), because of the lack of sufficiently large window in the dynamics, we can at least say that our findings are compatible with the theoretical prediction, in agreement with the results from Ref.~\onlinecite{coslovich_ozawa_kob}.

\begin{figure}[th]
\includegraphics[width=0.45\textwidth,trim = {0 0cm 0 0.1cm},clip]{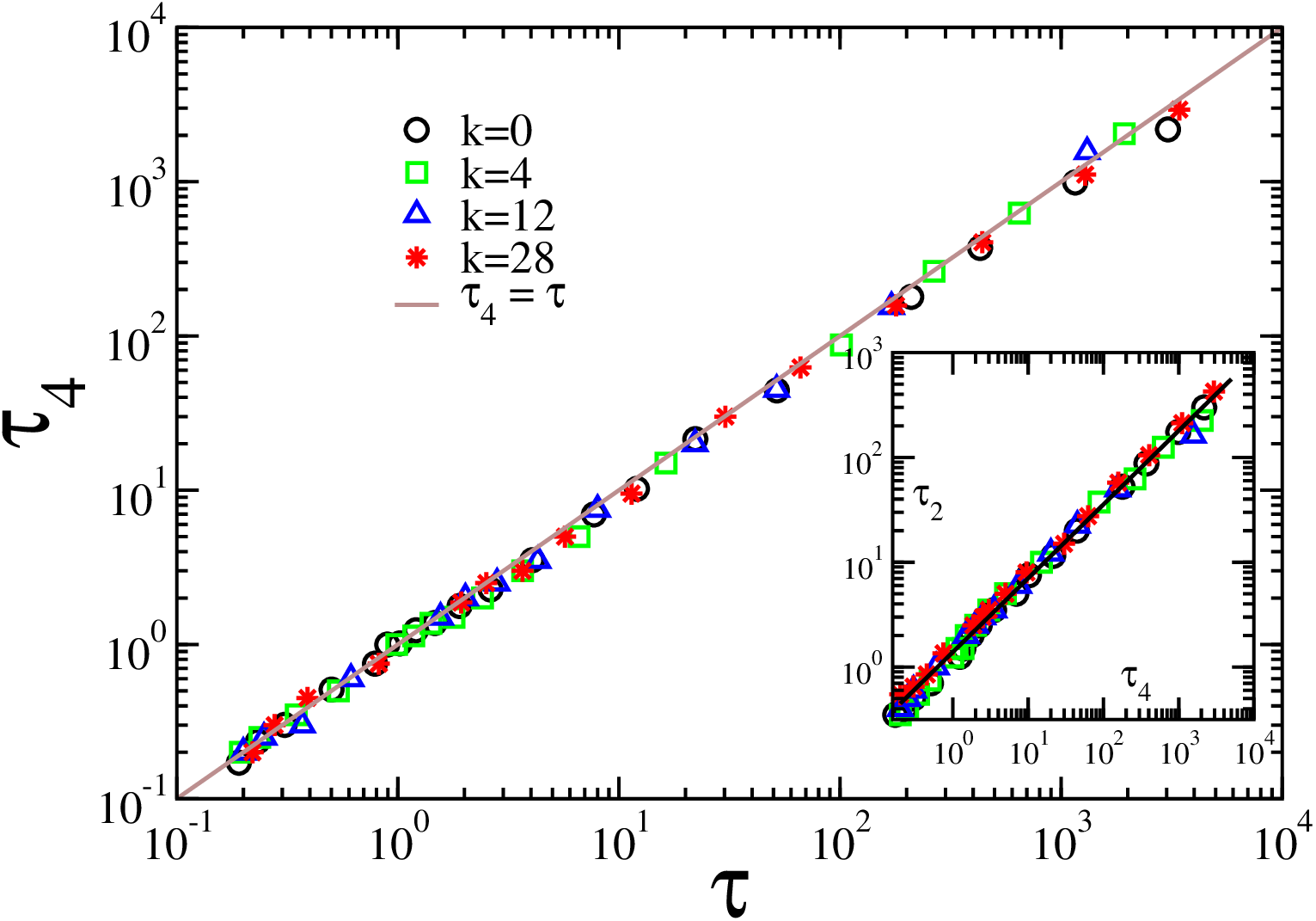}
\caption{The location of the peak in $\chi_4(t)$ as a function of the $\alpha$-relaxation time $\tau$. The symbols are for different values of $k$ and different $T$, and the solid line is a power law with exponent 1.0. 
Inset: $\tau_2$ as a function of $\tau_4$ showing a power law connection between the two quantities. The straight line has a slope of 0.70.\\}

\label{fig19_tau4_tau}
\end{figure}

Finally, we note that for large $\tau$ we find clear deviations of our data from the predicted power law in that the growth of $\chi_4^p$ is weaker than predicted. So in this regime, we can again invoke the argument that hopping processes decrease the cooperativity of the relaxation dynamics. 

Fig.~\ref{fig16_chi4_vs_T}(a) shows that the location of the peak in $\chi_4(t)$, $\tau_4$, quickly moves to larger times if the temperature is lowered. To determine the connection between the $\alpha$-relaxation time $\tau$ and the time scale $\tau_4$ we plot in Fig.~\ref{fig19_tau4_tau} $\tau_4$ as a function of $\tau$. Also included in the graph is the line $\tau_4=\tau$ (solid line) and one recognizes that all the data points fall on this line with high accuracy. Hence we can conclude that the time scale at which the system shows maximum cooperativity is on the time scale of the $\alpha$-process, which is in agreement with earlier results \cite{sastry_jack_star}. Also note that this conclusion is independent of $k$, i.e.~the strength of the mean-field character does not play a role for this result.
This result demonstrates that the $\alpha$-relaxation process is tightly related to the presence of the dynamical heterogeneities and that hence it is useful to study the latter in order to understand the slowing down of the relaxation dynamics. Finally we mention that the direct proportionality of $\tau_4$ to $\tau$ and the power law connection between $\tau_2$ and $\tau$, (see Fig.~\ref{fig15_tau2_vs_tau}) implies that we have the simple connection $\tau_2 \propto \tau_4^\kappa$, with an exponent $\kappa$ given by $b/(a+b)$, see Eq.~(\ref{eq31}). That this relation works indeed well is shown in the inset of Fig.~\ref{fig19_tau4_tau}. Since the exponent $\kappa$ is less than unity, we see that $\tau_2$ is smaller than $\tau_4$, as expected \cite{szamel-pre}. This can also be concluded from Fig.~\ref{fig9_msd_slope} where we have added in panels (a)-(c) the values of $\tau_4$ (downward arrows), in that one recognizes that at low $T$, these are indeed to the right of the arrows presenting $\tau_2$. These graphs also show that, interestingly, the (logarithmic) slope of the MSD at $t=\tau_4$ is independent of $T$ but weakly dependent on $k$.

\section{Summary and conclusion}
\label{sec:IV}

We have introduced a simple glass-forming system which allows to tune in a smooth manner its mean-field character. This is achieved by introducing additional $k$ ``pseudo neighbors'' with which a particle can interact. These additional interactions are long-ranged and hence with increasing $k$, each particle becomes increasingly connected with the rest of the system. However, since we also keep the original interaction between nearest-neighbor particles, our model has the advantage of maintaining a liquid-like structure even in the mean-field limit, i.e.~the nearest neighbor distances are always of the order of the particle diameter, which is in contrast to other models that allow tuning their mean-field character~\cite{mari-kurchan}. 

We find that the structure of the system, as characterized by the radial distribution function or the static structure factor, remains unchanged with the addition of the pseudo neighbours, also this in contrast to previous models. Due to the way the model is set up, it is possible to analytically calculate all the static structural properties of the system from the knowledge of the $k=0$ system. This allows us to understand that the additional interactions give rise to an effective potential that increases with $k$, thus influencing the relevant temperature scale of the system.

Due to the presence of the pseudo neighbors, the relaxation dynamics shows a very strong dependence on $k$ in that the onset temperature as well as the critical temperature of mode-coupling theory increase with increasing $k$. However, once the relaxation times are expressed in terms of the critical temperature of MCT one finds only a mild $k-$dependence, indicating that for this class of systems $T_c$ is the most relevant parameter for the dynamics, at least in the $T-$range investigated here. We note that the range in temperature in which MCT seems to give a good description of the relaxation dynamics increases systematically with increasing $k$, thus indicating that in the mean-field limit, the theory becomes exact. This is also confirmed by the observation that the dynamical heterogeneities, characterized by the dynamic susceptibility $\chi_4(t)$, decrease with increasing $k$.

It is often believed that the fragility of the glass-former is directly related to the presence of dynamical heterogeneities (or more precisely to the value of the stretching parameter $\beta$ in the Kohlrausch-Williams-Watts function used to fit the time-correlation functions)~\cite{bohmer_jcp_1993,xia_woly_prl,niss_tarjus}. Since we find that the fragility of the system increases with $k$ while the dynamic heterogeneity decreases we conclude that there is no such (strict) connection between these two quantities, although we do not want to exclude the possibility that in practice there might be a certain correlation. This result is in qualitative agreement with the findings in earlier studies \cite{sastry-SE,jeppe_2007}. Sengupta {\it et.~al.}~have, e.g.,~reported that compared to a three-dimensional system, the corresponding four-dimensional system was less heterogeneous but more fragile \cite{sastry-SE}. This is also corroborated by experimental data analyzed by Dyre, which indicate that there is no direct connection between fragility and heterogeneity \cite{jeppe_2007}.

The possibility to tune the mean-field character of the system without changing the structure also allows elucidating the relation between the non-Gaussian parameter $\alpha_2(t)$ and $\chi_4(t)$. While previous studies have often considered both functions to be indicators for the dynamical heterogeneities, our analysis shows that this is not the case at all since their dependence on $k$ is very different. Therefore our work clearly shows that these two observables convey information that is very different, a conclusion that is in line with previous results that showed that the peak in $\alpha_2(t)$ has a temperature dependence which differs from the one of $\chi_4^p$~~\cite{sastry-SE}. Furthermore, we also recall that for the MK-model, Ref.~\onlinecite{mari-kurchan}, one finds that $\chi_4^p$ decreases with increasing mean-field character of the system, i.e.~the same behavior as we have found here, but that also the value of $\alpha_2^p$ decreases, while in our case we find that $\alpha_2^p$ is independent of $k$. 
Also in the case of the Gaussian core model, it was found that it's $\alpha_{2}(t)$ peak is lower than the one for the Kob-Andersen model, whereas the $\chi_{4}$ peak is much higher \cite{kuni-jcp,kuni-pre}. The authors of these papers justified this results by stating that $\alpha_{2}$ provides a measure of the degree of dynamic heterogeneity and thus its peak value should be lower for more mean-field like models and $\chi_{4}$ provides a measure of the size of the domains and systems which have larger domains should have higher value of $\chi_{4}$.
Although this interpretation might apply to the Gaussian core model, it is not in agreement for the system studied here and hence not general. This suggests that further studies are required to understand the exact information provided by $\chi_{4}$ and $\alpha_{2}$ and if these two quantities are indeed related to each other. 

Finally, we also note that the decrease of $\chi_4$ with increasing $k$ can be due to the fact that the fluctuations in the overlap function do indeed decrease, i.e.~the relaxation dynamics of the system 
becomes more homogeneous, as expected for a mean-field-like system. However, since with increasing $k$ the characteristic temperatures of the system also increase, the fluctuations should decrease.
So for the moment, it is not clear which one of the two mechanisms is the main cause for the decrease of $\chi_4^p$ that we observe in the present work.

In an earlier study involving different glass-formers evidence was given that the locally preferred structures (LPS) are connected to the dynamics only for systems which are not mean field like~\cite{coslovich_mean_field}. The ability of the present model to continuously tune the mean-field behaviour makes it thus an ideal system to check the validity of this observation. Since we find that with increasing number of pseudo neighbours the LPS remains unchanged whereas the dynamics slows down, this suggests that with an increase in the mean field nature the correlation between the LPS and the dynamics decreases, a result that corroborates the earlier findings from Ref.~\onlinecite{coslovich_mean_field}.

The range of $k$ that we were able to access in the present simulation is relatively modest since for larger $k$ the relaxation dynamics became too slow to equilibrate the system within a reasonable amount of computer time. It is, however, of interest to make an educated guess on what will happen if $k$ is increased further. Our analytical results for the structure, Fig.~\ref{fig5_gr_eff_fig}, shows that with increasing $k$ the main peak in 
the effective radial distribution function becomes very high. In this limit one can thus expect that the contribution from the pseudo neighbors will start to dominate the one from the real nearest neighbors and hence will make the system mean-field like. However, from the graph we recognize that this increase becomes strong only once $k$ is larger than $O(10^2)$, i.e., a value that is at present somewhat beyond the reach of standard computer simulations. It can be expected, however, that in the near future improved algorithms will allow to deal with this bottleneck. In that case our approach will thus allow to make more stringent investigations on how the properties of a normal three dimensional glass-former can be connected to the corresponding system in the mean field limit.

This summary clearly indicates that the details how the mean-field limit is approached are important and future studies are needed to clarify this point. Finally, we note that the approach we propose here on how the mean-field character is tuned can be applied to any system. Hence it will be interesting to study whether other types of interaction potentials, such as the Coulomb potentials used to describe oxide glass-formers, will give qualitatively the same behavior, or in other words, whether the approach to the mean-field limit depends on the nature of the local structure of the system.\\[4mm]

{\bf Acknowledgements}\\
W.~K. is member of the Institut universitaire de France.
S.~M.~B thanks SERB for funding. U.~K.~N thanks CSIR for his fellowship. The authors thank C.~Dasgupta, D.~Coslovich, M.~K.~Nandi, and M.~Sharma for discussions. U.~K.~N thanks S.~Sengupta, A.~Banerjee and Md.~Alamgir for help with the initial setup of the system.\\[4mm]

{\bf Availability of Data}\\
The data that support the findings of this study are available from the corresponding author upon reasonable request.

\section{References}

\end{document}